\documentclass[aps,prb,twocolumn,superscriptaddress]{revtex4-2}

\usepackage[utf8]{inputenc}
\usepackage{physics}
\usepackage{amsmath,amsfonts,amsthm,amssymb,mathrsfs,latexsym,paralist}
\usepackage{graphicx}
\usepackage[colorlinks,citecolor=green,urlcolor=blue,bookmarks=false,hypertexnames=true]{hyperref} 

\begin{document}

\title{Quantum-State Projectors on Grassmannian: Geometry, Holonomy, and Topology}
\author{Shin-Ming Huang}
\affiliation{Department of Physics, National Sun Yat-sen University, Kaohsiung 80424, Taiwan}
\affiliation{Center for Theoretical and Computational Physics, National Sun Yat-sen University, Kaohsiung 80424, Taiwan}
\affiliation{Physics Division, National Center for Theoretical Sciences, Taipei 10617, Taiwan}
\date{\today}

\begin{abstract}
An isolated group of $k$ bands in an $N$-level quantum system defines a rank-$k$ spectral projector and hence a map into the complex Grassmannian $\mathrm{Gr}(k,N)$. For a smooth gapped Hamiltonian, this projector is globally smooth and periodic even when band topology obstructs a globally smooth periodic, or symmetry-compatible, Bloch frame. We take the globally defined differential $\dd P$ as the central object: it is the tangent field of the Grassmannian map, removes unphysical rotations within the selected subspace, and retains the physical interband transition. The interband block of this tangent data simultaneously determines the quantum metric and Berry curvature; the associated horizontal generator produces finite Grassmannian motion, while wedge products of $\dd P$ enter topological forms. We construct a shortest path between two projectors in the ambient Grassmannian and show that the singular values of its horizontal generator block are the principal angles between the endpoint subspaces. This construction provides a piecewise-geodesic interpretation of discrete geometric phases. In a separate development, we derive a basis-independent expression for the determinant of a multiband Wilson loop from traces of powers of an ordered projector product, without decomposing a degenerate band multiplet into individual bands. Finally, the same tangent-vector calculus organizes Chern characters, chiral winding numbers, and the time-reversal $\mathbb Z_2$ index. The resulting framework unifies local quantum geometry, finite subspace distance, holonomy, and topology while avoiding global gauge fixing. 
\end{abstract}

\maketitle
\section{Introduction}
The geometry of quantum states provides a common language for adiabatic dynamics, polarization, optical response, and the topology of Bloch bands. Berry holonomy and its non-Abelian generalization describe parallel transport of an isolated state or band multiplet~\cite{Simon1983,berry1984,wilczek1984}. Locally, the quantum geometric tensor (QGT) combines the Berry curvature with the Fubini--Study metric, which measures the distinguishability of nearby states~\cite{provost1980,Ma2010}. These geometric quantities govern phenomena ranging from Wannier localization and superfluid weight to optical spectral weight and nonlinear response~\cite{Torma2023,OnishiFu2024,YuReview2025,Jiang2025}. Their global integrals and symmetry-constrained descendants give the Chern, winding, and $\mathbb Z_2$ invariants used to classify insulating phases~\cite{Schnyder2008,fu2006}.

Quantum geometry has also become experimentally accessible. Momentum-resolved measurements have reconstructed the QGT or the full quantum metric of Bloch electrons in crystalline solids, while photonic and plasmonic platforms provide complementary direct probes~\cite{Cuerda2024,Kang2025,Kim2025Science}. Particularly striking progress has come from nonlinear transport: quantum-metric contributions have been identified through nonlinear Hall and nonreciprocal responses in topological antiferromagnets~\cite{Gao2023,Wang2023}, and through electrically tunable nonlinear magnetoresistance in spin-momentum-locked interfaces and topological-insulator surface states~\cite{Sala2025Science,Sala2026NatMater}. In parallel, resonant optical response, shift current, and intrinsic nonlinear Hall effects have been organized in terms of quantum-geometric data~\cite{Ahn2022,Avdoshkin2025,Ulrich2026}. These advances motivate formulations that are gauge invariant, stable at band degeneracies, and practical for higher-order derivatives and response calculations.

For a group of $k$ occupied bands, an eigenvector description carries a local $\mathrm{U}(k)$ gauge freedom. Berry connections and overlap matrices consequently depend on gauge. In a Chern band, no globally smooth periodic Bloch eigenbasis exists; in a time-reversal topological insulator, the obstruction concerns a globally smooth basis that also respects the time-reversal constraint~\cite{panati2007triviality,brouder2007,fu2006,bradlyn2017}. By contrast, if a smooth Bloch Hamiltonian retains a gap separating the selected band group, its spectral projector
$P(\vb{k})= \sum_{\alpha=1}^k u_\alpha(\vb{k}) u_\alpha(\vb{k})^\dagger$ is globally smooth and periodic. 
The topology is then carried by the map $P(\vb{k})$, rather than by singularities of a chosen frame. Projector formulas have long been central to Chern characters~\cite{avron1983homotopy,bellissard1994} and real-space formulations applicable in the presence of disorder~\cite{Lin2023,Shiina2025}; more recently, eigenprojector and multistate projector frameworks have been developed for degenerate-band geometry, nonlinear optical response, and crystalline observables~\cite{graf2021,MeraMitscherling2022,Avdoshkin2023,Avdoshkin2024,Avdoshkin2025,mitscherling2025,Oancea2026}. 
A rank-$k$ projector in an $N$-dimensional Hilbert space is a point of the complex Grassmannian $\mathrm{Gr}(k,N)=\mathrm{U}(N)/[\mathrm{U}(k)\times\mathrm{U}(N-k)]$~\cite{HuangGiataganas2026}. A band structure is therefore a map from the Brillouin torus to a Grassmannian. Recent work has also used the Grassmannian and its Pl\"ucker embedding~\cite{bouhon2023} to describe multiband generalized Landau levels~\cite{Liu2025GLL}.

The differential $\dd P$ is the first-order datum of this Grassmannian map. It discards changes of basis within either subspace and records only interband transitions.  It can give the quantum metric, Berry curvature, and higher geometric tensors. Tangent-space constructions are standard in the mathematical geometry of Grassmannians~\cite{Bendokat2024}; in quantum physics, projector differentials have appeared explicitly in K\"ahler descriptions of Chern bands, gauge-invariant projector calculus, and sub-bundle geometry~\cite{Mera2021_Kahler,Mera2021_engineering,mitscherling2025,Oancea2026}. 
Analytic formulas for first and second derivatives of eigenprojectors were developed from polynomial representations of the Hamiltonian, including projectors onto degenerate band multiplets; these formulas enabled a gauge-invariant construction of Christoffel symbols and Riemann curvature and a generalized Gauss--Bonnet analysis of two-band quantum-state manifolds~\cite{huang2025gaussbonnet}. The present study builds on this local and global projector calculus to connect tangent geometry with finite-distance geodesics, discrete holonomy, Wilson loops, and topology.

Here we develop a unified formulation from the Lie algebra of $\mathrm{U}(N)$. The Cartan decomposition separates stabilizer generators from physical, block-off-diagonal generators. Commutation with $P$ maps the latter isometrically to Grassmannian tangent vectors. Thus $\dd P$ has two complementary roles: physically, it characterizes interband changes of the quantum subspace; geometrically, it is the tangent movement on the Grassmann manifold. Its traces and wedge products give gauge-invariant metric, curvature, and topological data. We investigate how the tangent vectors enter the geodesic between two projectors. We also examine how a geometric phase emerges from the singular-value decomposition of projectors' overlap. 
We interpret Bargmann invariants as holonomies of piecewise-geodesic polygons and show that subdividing a geodesic edge does not change the associated phase. 
For multiband Wilson loops, we demonstrate that the determinant can be obtained solely from traces of powers of the ordered projector product. This result remains well defined at symmetry-enforced internal band degeneracies. Finally, differential forms built from $P$ organize even-dimensional Chern characters, while symmetry-adapted constructions yield odd-dimensional chiral winding numbers and the time-reversal $\mathbb Z_2$ index. Taken together, these results establish a gauge-invariant route from local quantum geometry to global topology without requiring a globally fixed Bloch frame.

The paper is organized as follows. Section~\ref{sec:level1} introduces spectral projectors and the Grassmannian, followed by the required unitary-group geometry. Section~\ref{sec:grass} develops tangent vectors, the quantum geometric tensor, geodesics, and discrete holonomy. Section~\ref{sec:diff_form} gives the differential-form formulation, and Sec.~\ref{sec:topology} applies it to Chern, chiral, and time-reversal invariants. Technical derivations are collected in the Appendices.

\section{Projectors and the Grassmannian} \label{sec:level1}
Suppose that a smooth Bloch Hamiltonian has an isolated group of $k$ occupied bands. Locally in momentum space, choose normalized eigenvectors and assemble them into the complete basis $\widetilde{\Psi}=\begin{pmatrix}u_1&u_2&\dots&u_N\end{pmatrix}$. We collect the occupied states in the $N\times k$ matrix $\Psi$ and the remaining $N-k$ states in $\Psi_\perp$, so that $\widetilde{\Psi}=(\Psi,\Psi_\perp)$. 
Any local unitary transformation $\Psi(\mathbf k) \rightarrow \Psi(\mathbf k)U(\mathbf k)$, represents the same occupied subspace. 
In condensed-matter physics this freedom is known as the non-Abelian Bloch gauge freedom, whereas in differential geometry it is viewed as a change of local orthonormal frame. 

The orthogonal projector onto the occupied subspace is
\begin{equation}
    P = \sum_{\alpha=1}^k u_\alpha u_\alpha^\dagger = \Psi \Psi^\dagger , \label{P}
\end{equation}
and $Q=I_N-P=\Psi_\perp\Psi_\perp^\dagger$ projects onto its orthogonal complement. Although the frames $\Psi$ and $\Psi_\perp$ may require multiple gauge patches, the spectral gap makes $P(\vb{k})$ and $Q(\vb{k})$ globally smooth and periodic whenever $H(\vb{k})$ is smooth and periodic in the Brillouin zone (BZ). By introducing the reference projector $P^{(0)}=\mqty(I_k & 0 \\ 0 & 0)$, where $I_k$ is the $k\times k$ identity matrix (the dimensions of the zero blocks will not be explicitly specified hereafter), the projector $P$ is written locally as
\begin{equation}
    P = \widetilde{\Psi} P^{(0)} \widetilde{\Psi}^\dagger . \label{P0}
\end{equation}
The projector is Hermitian ($P^\dagger = P$), idempotent ($P^2=P$), and has rank $k=\trace P$. 
Crucially, $P$ is invariant under independent gauge transformations (changes of frame) within the occupied and unoccupied subspaces,
\begin{equation}
    \widetilde{\Psi} \mapsto \widetilde{\Psi} \mqty(U_k & 0 \\ 0 & U_{N-k}), 
    \label{gauge_trans}
\end{equation}
where $U_k \in \mathrm{U}(k)$ and $U_{N-k} \in \mathrm{U}(N-k)$. These transformations form the stabilizer of $P$, encoding the gauge redundancy. The orbit of $P^{(0)}$ under $\mathrm{U}(N)$ is the complex Grassmannian
\begin{equation}
 \mathrm{Gr}(k,N)\simeq
 \frac{\mathrm{U}(N)}{\mathrm{U}(k)\times\mathrm{U}(N-k)},
\end{equation}
the manifold of $k$-dimensional complex subspaces of $\mathbb{C}^N$. Locally, this space is homeomorphic to a Euclidean space of real dimension $2k(N-k)$, as is also evident from the independent entries of the $(N-k)\times k$ complex off-diagonal block.

\section{Unitary-group geometry}
To understand the geometry of the Grassmannian, which is a quotient of the unitary 
group, we first recall the basic differential-geometric structure of $\mathrm{U}(N)$.

\subsection{Tangent space and adjoint representation}
Throughout this paper the group $G=\mathrm{U}(N)$ acts on the ambient Hilbert space. 
Its Lie algebra is the tangent space at the identity and consists of skew-Hermitian matrices. Indeed, for a smooth curve passing through the identity of $G$, it can be described by $\gamma(t)=e^{tX}$, with $X^\dagger=-X$, that satisfies $\gamma(0)=I_N$, $\gamma(t)^\dagger\gamma(t)=I_N$, and $\eval{\dv{}{t}\gamma(t)}_{t=0} =X$. Hence
\begin{equation}
\mathfrak{g}=\mathfrak{u}(N)=\{X\in M_{N\times N}(\mathbb{C}) \mid X^\dagger=-X \}.
\end{equation}
Geometrically, the Lie algebra is precisely the tangent space of the group manifold at the identity,
\begin{equation}
\mathfrak{g}=T_I G.
\end{equation}
Since $T_I G$ is a vector space, the addition of tangent vectors corresponds to the first-order approximation of group multiplication near the identity. The noncommutativity of the group appears through the Lie bracket at second order.

The Lie algebra defined at the identity uniquely determines the tangent space at any other point on the manifold. Specifically, at a given point $g \in G$, the tangent space $T_g G$ can be constructed via left translation:
\begin{equation}
T_g G = \{ gX \vert X \in \mathfrak{g} \}.
\end{equation}
The resulting vector fields are called \emph{left-invariant vector fields}. Unlike elements of the Lie algebra, tangent vectors in $T_gG$ are generally not skew-Hermitian; indeed, if $V=gX$, then $V^\dagger\neq -V$ in general. Nevertheless, because left translation creates an isomorphism $T_gG\simeq T_IG$, it is customary in physics to represent a tangent vector $V\in T_gG$ by its pullback $X=g^{-1}V\in\mathfrak{g}$, which is skew-Hermitian. In this way, tangent vectors at different points are regarded as elements of the same Lie algebra, and the distinction between $V$ and its Lie algebra representative $g^{-1}V$ is usually left implicit. We shall adopt this convention throughout this paper.

The adjoint representation maps $X\in\mathfrak{g}$ to the linear operator $\mathrm{ad}_X$, whose action on $Z\in\mathfrak{g}$ is
\begin{equation}
    \mathrm{ad}_X(Z)=\mathrm{ad}_X Z=\comm{X}{Z},
\end{equation}
where $\comm{X}{Z}=XZ-ZX$. By virtue of the Jacobi identity,
\begin{equation}
    [X,[Y,Z]]+[Y,[Z,X]]+[Z,[X,Y]]=0,
\end{equation}
the map $\mathrm{ad}$ constitutes a Lie algebra homomorphism that preserves the Lie bracket, satisfying
$[\mathrm{ad}_X,\mathrm{ad}_Y]=\mathrm{ad}_{[X,Y]}$. Evaluated on an arbitrary element $Z$, this homomorphism condition reads explicitly as
\begin{equation}
    [\mathrm{ad}_X,\mathrm{ad}_Y](Z) =\mathrm{ad}_{[X,Y]} (Z). \label{jacobi}
\end{equation}
The adjoint action also gives the unitary transformation of an operator $O$. For $g=e^X$,
\begin{align}
    \begin{split}
        O'&=e^{X} O e^{-X}=O + [X,O]+\frac{1}{2!} [X,[X,O]]+\cdots \\
        &= \sum_{k=0}^\infty \frac{1}{k!}(\mathrm{ad}_X)^k(O) = e^{\mathrm{ad}_X}O. \label{BCH}
    \end{split}
\end{align}
We will apply this identity to the projector in Eq.~(\ref{P0}).

\subsection{Lie brackets and curvature}
We equip $\mathrm{U}(N)$ with the bi-invariant Frobenius metric $\langle X,Y\rangle=\trace(X^\dagger Y)$ on skew-Hermitian generators. With this metric, $e^{tX}$ is both a one-parameter subgroup and a geodesic (straight line) through the identity. A small group-commutator loop obeys
\begin{align}
    \begin{split}
	\exp(-\epsilon Y) \exp(-\epsilon X) & \exp(\epsilon Y) \exp(\epsilon X) \\
	&= \exp \left( -\epsilon^2 [X, Y] + \mathcal{O}(\epsilon^3) \right) 
    \end{split}
\end{align}
for infinitesimal translations along the sequential directions $X$, $Y$, $-X$, and $-Y$. 
When $X$ and $Y$ do not commute, this loop fails to close at order $\epsilon^2$, measuring the noncommutativity of the infinitesimal motions. For left-invariant vector fields, the Levi-Civita connection of the bi-invariant metric is $\nabla_XY=\frac{1}{2}[X,Y]$.
The corresponding Riemann curvature tensor is defined as:
\begin{align}
    \begin{split}
        R(X,Y)Z&=(\nabla_X \nabla_Y - \nabla_Y \nabla_X- \nabla_{[X,Y]} ) Z \\
        &= -\frac{1}{4} [[X,Y],Z] .
    \end{split}
\end{align}
The sectional curvature of the plane spanned by $X$ and $Y$ is
\begin{align}
    \begin{split}
        K(X,Y)&=\frac{\langle X,R(X,Y)Y\rangle}{\langle X,X\rangle\langle Y,Y\rangle-\langle X,Y\rangle^2} \\
        &= \frac{1}{4}\norm{[X,Y]}^2
        \quad\text{for orthonormal }X,Y.
    \end{split}
\end{align}
Thus noncommuting tangent directions produce positive sectional curvature with the convention used here~\cite{lee2018introduction}.

\section{Grassmannian} \label{sec:grass}
\subsection{Tangent space and Cartan decomposition}
Because the Grassmannian is a quotient of $G$ under the equivalence relation, its tangent space is obtained by removing the stabilizer directions from $\mathfrak{g}$. Using Eq.~(\ref{BCH}), a curve through $P$ generated by $X\in\mathfrak{g}$ is
\begin{equation}
    P(t) = e^{Xt}P e^{-Xt}= e^{t \, \mathrm{ad}_X}P. \label{Pt}
\end{equation}
The derivative $\dot P(0)=[X,P]$ is the tangent vector at $P$, and the tangent space of the Grassmannian at $P$ is 
\begin{equation}
    T_P \mathrm{Gr}(k,N) =\{ \comm{X}{P} \mid X \in \mathfrak{u}(N)\} .
    \label{T_p}
\end{equation}
The tangent vector is not itself an element of the Lie algebra; it is the image of a Lie-algebra generator under the infinitesimal adjoint action, $T=\mathrm{ad}_X P$.
Although the Grassmannian itself is not a vector space, $T_P\mathrm{Gr}(k,N)$ is a real vector space of Hermitian matrices. The map $X\mapsto[X,P]$ has a kernel: generators that commute with $P$ do not move the Grassmannian point. Therefore, we classify the Lie-algebra generators by the Cartan decomposition.

The Lie algebra naturally separates into gauge (vertical) generators and physical (horizontal) generators:
\begin{equation}
    \mathfrak{g} = \mathfrak{k} \oplus \mathfrak{p},
\end{equation}
satisfies the commutator relations
\begin{equation}
    [\mathfrak{k},\mathfrak{k}] \subseteq \mathfrak{k},\quad
    [\mathfrak{k},\mathfrak{p}] \subseteq \mathfrak{p},\quad
    [\mathfrak{p},\mathfrak{p}] \subseteq \mathfrak{k}.
\end{equation}
For a gauge generator $A$, it commutes with $P$, $[A,P]=0$. These generators span the stabilizer subalgebra $\mathfrak{k}$; Eq.~(\ref{jacobi}) ensures that $[A,A']\in\mathfrak{k}$ whenever $A,A'\in\mathfrak{k}$. Its exponential is the stabilizer group $K\simeq\mathrm{U}(k)\times\mathrm{U}(N-k)$. 
On the other hand, a physical generator $X$ belongs to $\mathfrak{p}$, the orthogonal complement of $\mathfrak{k}$ with respect to the Frobenius metric. 
In differential geometry and gauge theory, the motions by $\mathfrak{k}$ and $\mathfrak{p}$ are called vertical and horizontal directions, respectively; the vertical generators correspond to pure gauge rotations within the occupied or unoccupied subspaces and therefore leave the projector unchanged, while the horizontal generators mix occupied and unoccupied states and generate genuine motion on the Grassmann manifold. 
In a block-matrix representation adapted to the occupied/unoccupied decomposition, 
elements of $\mathfrak{k}$ are block-diagonal and elements of $\mathfrak{p}$ are 
block-off-diagonal. 

For coordinates $k^\mu$ on parameter space, the tangent vectors are
\begin{equation}
    T_{\mu} = \partial_{\mu}P = [X_\mu, P], \label{Tmu0}
\end{equation}
where $\partial_\mu:=\partial/\partial k^\mu$. 
A fundamental property of the tangent vector follows from differentiating $P^2=P$: one obtains $PT_\mu P=QT_\mu Q=0$, and hence
\begin{equation}
    T_{\mu} = QT_{\mu}P + P T_{\mu}Q. \label{Tmu}
\end{equation}
The tangent vector therefore connects the occupied and unoccupied subspaces. This makes its physical content precise: it is insensitive to gauge transformations  
within either subspace, whereas $QT_\mu P$ and its adjoint encode their relative interband change. Norms, traces, and ordered products of these tangent vectors are gauge invariant and supply the geometric and topological quantities developed below.

The unique horizontal generator associated with $T_\mu$ is
\begin{equation}
    X_{\mu} = \comm{T_\mu}{P}. \label{Xmu}
\end{equation}
By this definition, the translation generator has the same feature as the tangent vector in Eq.~\eqref{Tmu}:
\begin{equation}
    X_{\mu} = QX_{\mu}P + P X_{\mu}Q .
\end{equation}
Indeed, $[[T_\mu,P],P]=T_\mu$. Thus we can define the map 
\begin{equation}
	\pi_P(\cdot)=[~\cdot~,P], \label{pi_P}
\end{equation}
with $\pi_P^2=1$, that maps mutually between Hermitian tangent vectors and horizontal skew-Hermitian generators. 
With the inner products used below, it is also an isometry. 

In the eigenbasis $\widetilde{\Psi}$, the tangent vector and the translation generator take 
the block-off-diagonal form
\begin{equation}
    T_\mu^{(0)} = \widetilde{\Psi}^\dagger T_\mu \widetilde{\Psi} 
    = \mqty(0 & B_\mu^{\dagger} \\ B_\mu & 0), \label{Tt}
\end{equation}
where $B_{\mu} \in M_{(N-k) \times k}(\mathbb{C}) $ is the interband transition matrix. 
This implies
\begin{equation}
    T_\mu = \Psi_\perp B_\mu \Psi^\dagger + \Psi  B_\mu^\dagger \Psi_\perp^\dagger .
\end{equation}
Similarly, 
\begin{equation}
    X_{\mu}^{(0)} = \widetilde{\Psi}^\dagger X_\mu \widetilde{\Psi} 
    = \mqty(0 & -B_{\mu}^{\dagger} \\ B_{\mu} & 0), \label{X0}
\end{equation}
In this basis the stabilizer generators are $\mqty(A_\mu&0\\0&C_\mu)$, where $A_\mu$ and $C_\mu$ are skew-Hermitian matrices of sizes $k\times k$ and $(N-k)\times(N-k)$, respectively. Appendix~\ref{appendix:generator} relates the horizontal generator to derivatives of the Hamiltonian.

\subsection{Quantum geometric tensor}

We endow the Grassmannian with the metric induced by its standard embedding in the Hermitian matrices:
\begin{equation}
    g_{\mu \nu  } = \expval{T_{\mu},T_{\nu}} :=\trace \left( T_{\mu} ^\dagger T_{\nu} \right), \label{metric}
\end{equation}
where $T_{\mu}$’s are coordinate tangent vectors. 
This convention is twice the one often used for the quantum metric: $g_{\mu\nu}=2\Re\sum_{\alpha\in\mathrm{occ.}}\matrixel{\partial_\mu u_\alpha}{Q}{\partial_\nu u_\alpha}$. In the eigenbasis,
\begin{equation}
    g_{\mu \nu} = 2\,\mathrm{Re}\,\mathrm{tr}(B_{\mu}^{\dagger} B_{\nu}).
    \label{gmn3}
\end{equation}
Because $\pi_P$ is an isometry, the same metric is $g_{\mu\nu}=\expval{X_{\mu},X_{\nu}}  = -\trace(X_\mu X_\nu)$, despite $T_\mu$ being Hermitian and $X_\mu$ skew-Hermitian.

As shown in Sec.~\ref{subsec:2form}, the trace Berry curvature in the same convention is
\begin{equation}
    F_{\mu \nu} = i \trace \left( P \comm{T_\mu}{T_\nu} \right) = -2 \Im \trace \left( B_\mu^\dagger B_\nu\right).
 \end{equation}
The quantum geometric tensor (QGT) $Q_{\mu \nu}$ unifies the metric and the Berry curvature into a single Hermitian object: 
\begin{equation}
    Q_{\mu \nu} = \mathrm{tr}(B_\mu^\dagger B_\nu) 
    = \frac{1}{2}(g_{\mu\nu} - i F_{\mu\nu}), \label{QGT}
\end{equation}
Thus $g_{\mu\nu}=2\Re Q_{\mu\nu}$ and $F_{\mu\nu}=-2\Im Q_{\mu\nu}$. Positivity of $Q_{\mu\nu}$ constrains the metric and curvature and yields familiar quantum-geometric inequalities~\cite{Ozawa2021,Mera2021_Kahler,Yu2025Wilson}.

For a two-band model (one occupied band), $B_\mu= \theta_\mu e^{i \phi_\mu}$ ($\theta_\mu \ge 0$) is a scalar complex number, and the QGT becomes the $2 \times 2$ matrix
\begin{equation}
    \hat{Q}=\mqty(\theta_1^2 & \theta_1 \theta_2 e^{-i \phi} \\ \theta_1 \theta_2 e^{i \phi} & \theta_2^2),
\end{equation}
where $\phi = \phi_1 - \phi_2$. 
With the metric convention of Eq.~(\ref{metric}), one finds $\sqrt{\det g}=|F_{12}|=2|\theta_1\theta_2\sin\phi|$~\cite{wang2021}. Thus the Berry curvature is the oriented area density of the two-band image, while the $2\times2$ QGT has one vanishing eigenvalue. In two-dimensional two-band models, zeros of $\det g$ can form curves across which the orientation, and hence the sign of the Berry curvature, changes~\cite{Ozawa2021,Mera2021_Kahler,wang2021,huang2025gaussbonnet}. On such a singular curve the differential of the map loses rank: there exists a nonzero parameter-space direction $v^\mu$ for which $v^\mu\partial_\mu P=0$.

\subsection{Geodesic}
Consider a smooth curve $P(t)=P(\vb{k}(t))$ on the Grassmann manifold. The tangent vector to the curve is $\dot P = \dot k^\mu \mathbf e_\mu$, where $\mathbf e_\mu=\partial_\mu P$ is identical to the tangent basis $T_\mu$ introduced previously. 
Throughout this work, the Einstein summation convention is assumed. The Grassmann manifold is endowed with a Riemannian metric $g_{\mu \nu} = \trace(\mathbf e_\mu \mathbf e_\nu)$, which defines the infinitesimal distance between neighboring points. 
The length of the curve parametrized by $\vb{k}(t)$ is therefore  
\begin{equation}
    l = \int_{0}^{1} \sqrt{\trace (\dot P^2)}\,\dd t
    = \int_{0}^{1} \sqrt{g_{\mu \nu}(\vb{k})\dot k^\mu\dot k^\nu}\,\dd t .
\end{equation}
The stationary curves are obtained by varying this functional. The resulting geodesic equation takes the standard form~\cite{lee2018introduction}
\begin{equation}
    \dv[2]{k^\rho}{t} + {\Gamma^\rho}_{\kappa \lambda}\dv{k^\kappa}{t} \dv{k^\lambda}{t} = 0,
\end{equation}
where ${\Gamma^\rho}_{\kappa\lambda}$ are the Christoffel symbols associated with the metric $g_{\mu \nu}$. In general, finding the geodesic connecting two prescribed points in $\vb{k}$-space requires numerical integration~\cite{Chen_2025}. The Christoffel symbols are the coordinate representation of the Levi-Civita connection. Specifically, they are defined by 
\begin{equation}
	\nabla_{\mu}\mathbf e_\nu = {\Gamma^\alpha}_{\mu\nu}\mathbf e_\alpha.
\end{equation}
Analytic formulas for the first and second derivatives of individual and degenerate-band projectors can be used to compute the associated Christoffel symbols and Riemann curvature directly from the Hamiltonian. This derivative calculus was developed and applied to a generalized Gauss--Bonnet relation for two-band quantum-state manifolds in Ref.~\cite{huang2025gaussbonnet}.

With an affine parameter, the geodesic problem is equivalent to minimizing the action
\begin{equation}
    S = \frac{1}{2}\int_{0}^{1}\trace(\dot P^2)\,\dd t .
\end{equation}
For fixed endpoints, varying the action and integrating by parts gives
\begin{equation}
    \delta S = \int_{0}^{1} \trace (\dot P \delta \dot P) \dd t = -\int_{0}^{1} \trace (\ddot P \delta P) \dd t =0.
\end{equation}

When the curve is constrained to the physical image $ P (\mathcal M), ~ \mathcal M=\mathrm{BZ} $, its admissible variations take the form $ \delta P=\partial_\mu P\,\delta k^\mu$. 
Stationarity then gives $ \trace(\ddot P\,\partial_\mu P)=0$, 
which is equivalent to the above Christoffel geodesic equation in parameter space. 

Now we consider a less constrained problem: the shortest path between \(P(0)\) and \(P(1)\) in the full ambient Grassmannian. 
In this problem, $P(t)$ is allowed to explore all rank-$k$ projectors. The constraint $P^2=P$ restricts the variation to $\delta P=P\delta P Q+Q\delta P P$. 
Stationarity with respect to every Grassmannian tangent variation gives 
\begin{equation}
    \ddot P = P \ddot P P +Q \ddot P Q.  \label{accel_P}
\end{equation}
Thus, $[\ddot P,P]=0$, and also the acceleration is normal to the tangent space: $\tr(\ddot P \dot P)=0$. This leads to 
\begin{equation}
 \dv{t} \tr(\dot P^2)=0.
\end{equation}
The resulting ambient geodesic need not lie in the physical image \(P(\mathcal M)\). In what follows, we focus on the geodesic problem in the full ambient Grassmannian; whether this path lies in \(P(\mathcal M)\) is model dependent.

Define the unique horizontal generator associated with the tangent vector by $X=[\dot P,P]$. 
Differentiating this expression yields
\begin{equation}
	\dot X= [\ddot P,P]+[\dot P,\dot P]=[\ddot P,P]=0, 	
\end{equation}
so $X$ is constant along an ambient Grassmannian geodesic. 

From the orbit $P(t)=e^{\tilde{X} t}P(0)e^{-\tilde{X}t}$, where $\tilde{X}=X+A$ and $X \in\mathfrak{p}$ and $A\in\mathfrak{k}$, the acceleration at \(t=0\) is
\begin{align*}
	\begin{split}
	\ddot P(0)&=[X+A,[X+A,P(0)]] \\
	&=[X,[X,P(0)]]+[A,[X,P(0)]] \\
	& =[X,[X,P(0)]]+[[A,X],P(0)],
	\end{split}
\end{align*}
in the last equality where we use the Jacobi identity and $[A,P(0)]=0$. In the last equality, 
the first term is block diagonal and normal to the Grassmannian, whereas the second is block off-diagonal and tangent to the Grassmannian. 
However, Eq. (\ref{accel_P}) requires $\ddot P(0)$ normal to the Grassmannian. Therefore, $[[A,X],P(0)]=0$. 
Since $[A,X] \in \mathfrak p$, applying the map $ \pi_{P(0)}$ defined in Eq.~(\ref{pi_P}) twice, it answers that 
\begin{equation}
	[A,X]=0.
\end{equation}
Consequently, the geodesic curve reads 
\begin{equation}
	P(t)=e^{X t} e^{A t}P(0)e^{-A t}e^{-Xt} = e^{X t} P(0) e^{-Xt},
\end{equation}
as stated in Eq. (\ref{Pt}). 
Thus, although the commuting vertical component is not uniquely determined, it has no effect on the projector trajectory. We may therefore set $A=0$, so that the basis follows the geodesic without any additional rotation within the occupied or unoccupied subspaces. This is the usual parallel-transport choice~\cite{berry1989quantum,Resta2000}.

\subsection{Horizontal generator}
We now determine the horizontal generator $X$ connecting $P_1=\Psi_1\Psi_1^\dagger$ and $P_2=\Psi_2\Psi_2^\dagger$. The singular-value decomposition (SVD)
\begin{equation}
    \Psi_1^\dagger \Psi_2 = \mathcal{R}_1 \Sigma \mathcal{R}_2^\dagger,
    \label{Psi1Psi2}
\end{equation} 
defines the principal angles $\theta_i\in[0,\pi/2]$ through $s_i=\cos\theta_i$, where $\Sigma=\mathrm{diag}(s_1,\ldots,s_k)$. The gauge transformations 
$\Phi_i=\Psi_i\mathcal{R}_i$ leave $P_i$ invariant and give $\Phi_1^\dagger\Phi_2=\Sigma$. Consequently,
\begin{equation}
P_1 P_2 = \Phi_1 \Sigma \Phi_2^\dagger.  \label{P1P2}
\end{equation}
The SVD determines the principal frames associated with the two $k$-planes. A detailed geometric interpretation of principal angles can be found in Ref.~\cite{HuangGiataganas2026}.

For completeness, we first recall a logarithmic expression for the horizontal generator~\cite{batzies_2015}. 
By noting that
\begin{equation}
    e^X P_1 - P_1 e^{-X}=\sinh(X),
\end{equation}
where $\cosh X = P_1 \cosh (X)P_1 + Q_1 \cosh (X) Q_1$ and $\sinh X = P_1 \sinh (X)Q_1 + Q_1\sinh (X) P_1$ (with $Q_{i} = I_N -P_{i}$) are considered, $P_2$ can be expressed as:
\begin{align}
    \begin{split}
        P_2 & = e^X P_1 e^{-X} = \left( \sinh(X)+P_1 e^{-X}\right)e^{-X} \\
        &=\frac{1}{2}I_N - \frac{1}{2}(I_N -2P_1)e^{-2X}.
    \end{split}
\end{align}
Using $(I_N-2P_i)^2=I_N$, one obtains
\begin{equation}
    X = \frac{1}{2} \log[(I_N - 2P_2)(I_N - 2P_1)]
\end{equation}
with a consistent branch of the matrix logarithm. The principal branch gives the minimal generator when all principal angles are smaller than $\pi/2$. If an angle equals $\pi/2$, the endpoint lies on the cut locus and the minimizing geodesic is not unique. 
This closed-form expression relies on a matrix logarithm, which is generally difficult to evaluate and interpret analytically. In what follows, we provide an alternative construction that explicitly reveals how the principal angles manifest within the generator.

For notational simplicity we take $k\le N-k$; the complementary case follows upon exchanging $P$ and $Q$. The compact SVD of the off-diagonal block in Eq.~(\ref{Tt}) is
\begin{equation}
    B=R_\perp \Theta R_\parallel^\dagger,
    \label{B}
\end{equation}
where $\Theta=\mathrm{diag}(\theta_1,\ldots,\theta_k)$ and $R_\parallel$, $R_\perp$ have orthonormal columns:
\begin{align}
    \begin{split}
        R_\parallel &= \mqty(r_\parallel^{(1)}, & \hdots, & r_\parallel^{(k)}) \in M_{k \times k}(\mathbb{C}), \\       
    R_\perp &= \mqty(r_\perp^{(1)}, & \hdots, & r_\perp^{(k)}) \in M_{(N-k) \times k}(\mathbb{C}),
    \end{split}
\end{align}
such that $R_\perp^\dagger R_\perp = R_\parallel^\dagger R_\parallel = I_{k}$. Choosing $\Theta$ as a square matrix avoids the need to distinguish between $\Sigma^\dagger \Sigma$ and $\Sigma \Sigma^\dagger$ in the subsequent derivations. The SVD determines the spectrum of $T^{(0)}$, which is symmetric about zero:
\begin{equation}
    T^{(0)} = \sum_{\alpha=1}^k \theta_\alpha \left( v_{+}^{(\alpha)}v_{+}^{(\alpha)\dagger}
    - v_{-}^{(\alpha)}v_{-}^{(\alpha)\dagger} \right).
\end{equation}
Consequently, the spectrum consists of $k$ positive eigenvalues ($\theta_\alpha \ge 0$), $k$ negative eigenvalues ($-\theta_\alpha \le 0$), and $N-2k$ vanishing eigenvalues. The corresponding eigenvectors are given by
\begin{equation}
    v_{+}^{(\alpha)} = \frac{1}{\sqrt{2}}\mqty(r_\parallel^{(\alpha)} \\ r_\perp^{(\alpha)}),~~
    v_{-}^{(\alpha)} = \frac{1}{\sqrt{2}} \mqty(~r_\parallel^{(\alpha)} \\ -r_\perp^{(\alpha)}).
\end{equation}

Using this spectral decomposition, $X^{(0)}$ and $T^{(0)}$ can be expressed compactly as
\begin{align}
    X^{(0)}=\mathscr{Z} \mqty(0 & -\Theta \\ \Theta  & 0) \mathscr{Z}^\dagger, ~~
    T^{(0)} =\mathscr{Z} \mqty(0 & \Theta \\ \Theta  & 0) \mathscr{Z}^\dagger, 
\end{align}
where $\mathscr{Z} = \mqty(R_\parallel & 0 \\ 0 & R_\perp)$ is an $N \times 2k $ matrix. We can then explicitly evaluate the time-evolved projector $P(t)$ using the adjoint action:
\begin{widetext}
\begin{align}
    \begin{split}
        P(t)&= e^{t \, \mathrm{ad}_X } P  = \widetilde{\Psi}_1  \left( e^{t \, \mathrm{ad}_{X^{(0)}} } P^{(0)} \right) \widetilde{\Psi}_1^\dagger\\
        & = \widetilde{\Psi}_1 \left[ P^{(0)} + \frac{1}{2} \mathscr{Z} \mqty((\cos(2\Theta t)-I_k) & \sin(2\Theta t) \\  \sin(2\Theta t)  & -(\cos(2\Theta t)-I_k))  \mathscr{Z}^\dagger \right] \widetilde{\Psi}_1^\dagger \\
        &= \widetilde{\Psi}_1 \left[\mqty(R_\parallel & 0 \\ 0 & R_\perp) 
        \mqty(\cos^2(\Theta t) & \sin(\Theta t) \cos(\Theta t) \\ \sin(\Theta t)\cos(\Theta t)  & \sin^2(\Theta t))  \mqty(R_\parallel^\dagger & 0 \\ 0 & R_\perp^\dagger) \right] \widetilde{\Psi}_1^\dagger\\
        &= \mqty(\Psi_1 & \Psi_{1\perp}) \mqty( R_\parallel \cos(\Theta t) \\ R_\perp \sin(\Theta t)) 
        \mqty( \cos(\Theta t) R_\parallel^\dagger  & \sin(\Theta t) R_\perp^\dagger) \mqty(\Psi_1^\dagger \\ \Psi_{1\perp}^\dagger)    \\
        &= \left(\Psi_1 R_\parallel \cos(\Theta t) + \Psi_{1\perp} R_\perp \sin(\Theta t) \right)  \left(\Psi_1 R_\parallel \cos(\Theta t) + \Psi_{1\perp} R_\perp \sin(\Theta t) \right)^\dagger
    \end{split}
\end{align}
\end{widetext}

At $t=1$, equality of the projectors implies equality of their frames up to a unitary matrix:
\begin{equation}
\Psi_2 = \left[ \Psi_1 R_\parallel \cos\Theta + \Psi_{1\perp} R_\perp \sin\Theta \right]U_k,
\end{equation}
Taking the inner product with $\Psi_1^\dagger$ gives $\Psi_1^\dagger\Psi_2=R_\parallel\cos\Theta\,U_k$. Comparison with Eq.~(\ref{Psi1Psi2}) permits the choices
\begin{equation}
R_\parallel = \mathcal{R}_1 \quad \text{and} \quad U_k = \mathcal{R}_2^\dagger ,
\end{equation}
which immediately gives
\begin{equation}
\cos \Theta = \Sigma.
\end{equation}
Thus the singular values of the geodesic generator block $B$ are the principal angles between the endpoint subspaces.
We can then rewrite Eq.~(\ref{Psi1Psi2}) in the form
\begin{equation}
\Psi_1^\dagger \Psi_2 = S_{12} \mathcal{R}_{12}, \label{P1P2_v2}
\end{equation}
where $S_{12}=\mathcal{R}_1\cos\Theta\,\mathcal{R}_1^\dagger$ is positive semidefinite and $\mathcal{R}_{12}=\mathcal{R}_1\mathcal{R}_2^\dagger$ is the unitary polar factor. The former records the reduction of overlap amplitudes, whereas products of the latter encode the holonomy around a closed polygon. Although $\mathcal{R}_{12}$ depends covariantly on the endpoint frames, the determinant of its product around a loop is gauge invariant.

Next, taking the inner product of the state with $\Psi_{1\perp}^\dagger$ yields
\begin{equation}
    \Psi_{1\perp}^\dagger \Psi_2 = R_\perp \sin(\Theta )\mathcal{R}_2^\dagger ,
    \label{P12perp}
\end{equation}
For nonzero $\sin\theta_\alpha$, this gives
\begin{equation}
    R_\perp = \Psi_{1\perp}^\dagger \Psi_2 \mathcal{R}_2 [\sin(\Theta)]^{-1}.
\end{equation}
Substituting this back into Eq.~(\ref{B}), the matrix $B$ can be reformulated as
\begin{equation}
    B = \Psi_{1\perp}^\dagger \Psi_2 \mathcal{R}_2 \frac{\Theta}{\sin \Theta}   \mathcal{R}_1^\dagger.
\end{equation}
Reconstructing the full generator via $X=\widetilde{\Psi} X^{(0)} \widetilde{\Psi}^\dagger=\Psi_\perp B \Psi_1^\dagger - \Psi_1 B^\dagger \Psi_\perp^\dagger$,
we obtain
\begin{align}
    \begin{split}
        X &= \Psi_{1\perp} \Psi_{1\perp}^\dagger \Psi_2 \mathcal{R}_2 \frac{\Theta}{\sin \Theta}  \mathcal{R}_1^\dagger \Psi_{1}^\dagger - \mathrm{h.c.}\\
        &= Q_1  \Phi_2 \frac{\Theta}{\sin \Theta}  \Phi_1^\dagger P_1 - \mathrm{h.c.}
    \end{split} \label{X_gen}
\end{align}
This constructs the generator from the SVD of the endpoint overlap in Eq.~(\ref{P1P2}). 
Note that if any principal angles vanish, $\sin \Theta$ is not invertible. For these specific modes, the corresponding columns of $R_\perp$ contribute neither to $\Psi_{1\perp}^\dagger \Psi_2$ in Eq.~(\ref{P12perp}) nor to $B$; we can therefore safely set these elements to the continuous limit $\lim_{\theta \rightarrow 0} \frac{\theta}{\sin \theta}=1$. 
When all principal angles satisfy $\theta_\alpha<\pi/2$, the shortest Grassmannian geodesic is unique. If at least one principal angle equals $\pi/2$, the unitary factor $\mathcal{R}_{12}$, and hence the corresponding horizontal generator in the SVD construction, not unique. Nevertheless, each admissible construction still produces a shortest geodesic. 

Finally, the geodesic distance is evaluated as
\begin{align}
    \begin{split}
        l &= \int_0^1 \sqrt{\trace(\dot{P}(t)^2)}dt \\
        &= \sqrt{-2 \trace (PX^2P)} \int_0^1  dt \\
        &= \sqrt{2 \trace(B^\dagger B)} \\
        & = \sqrt{2 \sum_{\alpha=1}^k \theta_\alpha^2} ,
    \end{split}
\end{align}
which is $\sqrt{2}$ times the Euclidean norm of the principal-angle vector. With the more common convention $g^{\rm conv}_{\mu\nu}=g_{\mu\nu}/2$, the Grassmannian distance is $\|\boldsymbol\theta\|_2$~\cite{batzies_2015,Bendokat2024,Avdoshkin2024}.

\subsection{Berry phase}
\subsubsection{Bargmann invariant} \label{subsec:bargmann}
Consider a discrete cyclic transition process from $P_1$ to $P_2$, then to $P_3$, and eventually returning to $P_1$, where $\Psi_j \equiv \Psi(\vb{k}_j)$ denotes the multiplet of states at each point. Throughout this loop, the system accumulates a geometric Berry phase $\varphi_B$ determined by
\begin{equation}
	\varphi_B = -\arg  \det(\mathcal{W}_{123}), \label{phi_B}
\end{equation}
where $\mathcal{W}_{123}=M_{1,2}M_{2,3}M_{3,1}$ is the overlap product, with
\begin{equation}
	(M_{i,j})_{\alpha,\beta} = (\Psi_i^\dagger\Psi_j)_{\alpha,\beta}=\braket{u_\alpha(\vb{k}_i)}{u_\beta(\vb{k}_j)}  \label{M_ij}
\end{equation}
($\alpha,\beta=1,\dots,k$). 
The minus sign in Eq.~\eqref{phi_B} appears because the Berry phase is the argument of $\mathcal{W}_{321}$ rather than that of $\mathcal{W}_{123}$.
Under independent gauge transformations at the three vertices, $\mathcal W_{123}$ transforms by conjugation at the base point, so its determinant is gauge invariant. The polar decomposition $\Psi_i^\dagger\Psi_j=S_{ij} \mathcal R_{ij}$ as in Eq. (\ref{P1P2_v2}) separates a positive-semidefinite amplitude factor from a unitary factor. Provided adjacent overlap matrices are nonsingular, the positive determinants do not affect the phase, and
\begin{equation}
    \varphi_B = -\Im \log \det \left( \mathcal{R}_{12 }\mathcal{R}_{23} \mathcal{R}_{31}\right).
    \end{equation}

A natural question arises: if the Berry phase is conventionally interpreted as the accumulated phase along an adiabatic path, what is the underlying curve connecting these three discrete points? We show below that the effective paths are precisely the geodesics connecting each pair of states. To verify this, we insert an intermediate point $P_t$ along the geodesic connecting $P_1$ and $P_2$, and examine whether the total phase remains invariant.
The corresponding Wilson loop structure requires evaluating the operator product $\Psi_1^\dagger P_t P_2 P_3 \Psi_1$. Assuming that $P_1$ and $P_2$ are connected by the geodesic generator $X$, the intermediate projector is parametrized as $P_t = e^{Xt} P_1 e^{-Xt}$, with the final endpoint given by $P_2 = e^{X} P_1 e^{-X}$. Utilizing the relation $\Psi_2 = e^X \Psi_1 \mathcal{R}_{12}$, the initial segment of the loop simplifies to:
\begin{align}
    \begin{split}
        \Psi_1^\dagger P_t P_2 &= \Psi_1^\dagger (e^{Xt} \Psi_1 \Psi_1^\dagger e^{-Xt} ) (e^{X}\Psi_1 \mathcal{R}_{12} ) \Psi_2^\dagger \\
        &= (\Psi_1^\dagger e^{Xt} \Psi_1)(\Psi_1^\dagger e^{X(1-t)} \Psi_1) \mathcal{R}_{12} \Psi_2^\dagger  \\
        & = (\mathcal{R}_1 \cos (\Theta t) \mathcal{R}_1^\dagger ) (\mathcal{R}_1 \cos (\Theta (1-t)) \mathcal{R}_1^\dagger )  \mathcal{R}_{12} \Psi_2^\dagger      \\
        & =S(t) \mathcal{R}_{12} \Psi_2^\dagger 
    \end{split}
\end{align}
where $S(t)=\mathcal{R}_1\cos(\Theta t)\cos[\Theta(1-t)]\mathcal{R}_1^\dagger$ is positive semidefinite for $0\leq\theta_\alpha<\pi/2$. It contributes no phase to the determinant. 
The phase acquired along the compound path $P_1 \to P_t \to P_2$ is therefore entirely encoded in $\mathcal{R}_{12}$, matching the direct $P_1 \to P_2$ path exactly. This reinforces the conclusion that no geometric phase is accumulated during parallel transport along a geodesic segment.

\begin{figure}[tbp]
    \centering
    \includegraphics[width=0.4\textwidth]{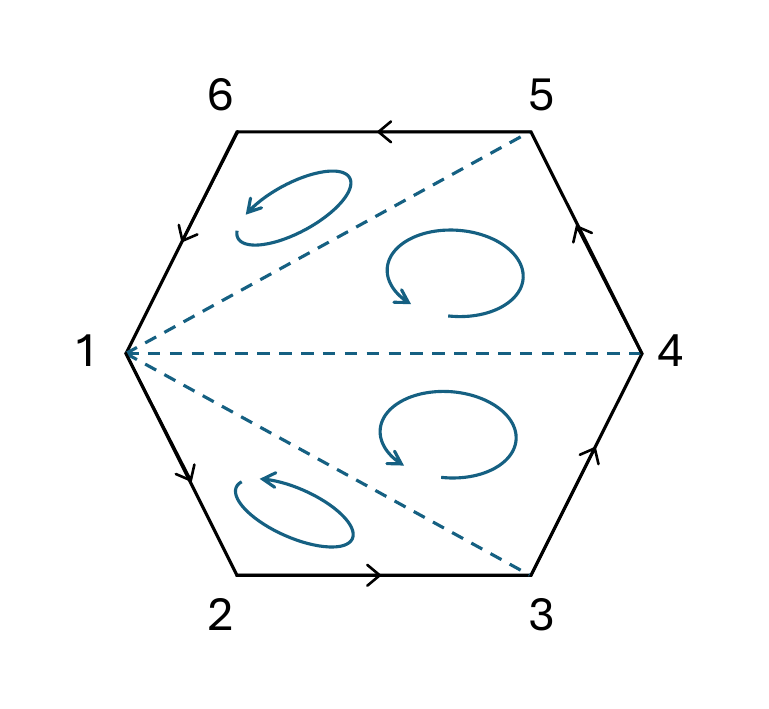}
    \caption{Triangulation of a six-point Bargmann invariant into four third-order invariants. Arrows indicate the orientation of each overlap product.}
    \label{fig:bargmann}
\end{figure}

For a single-state system, $P = \dyad{u}$, the Berry phase can be expressed by a product of projectors. For instance, 
Eq.~(\ref{phi_B}) becomes $\varphi_B=-\arg\trace(P_1P_2P_3)$. The trace is a third-order Bargmann invariant~\cite{bargmann1964,Simon1993,hassan2018,Avdoshkin2023,Zhang2025,mitscherling2025}; its $n$-point generalization is
\begin{equation}
 \Phi(1,2,\dots,n) = \trace(P_1 P_2 \dots P_n).
\end{equation}
For nonvanishing denominators, an $n$-point invariant can be triangulated into two- and three-point invariants~\cite{Avdoshkin2023}:
\begin{equation}
	\Phi(1,2,\dots,n) = \frac{\prod_{j=2}^{n-1}\Phi(1,j,j+1) }{\prod_{j=3}^{n-1} \Phi(1,j)}. \label{barg_decomp}
\end{equation}
For example, Fig.~\ref{fig:bargmann} shows the decomposition of a six-point invariant,
\begin{equation*}
	\Phi(1,2,3,4,5,6) = \frac{\Phi(1,2,3) \Phi(1,3,4) \Phi(1,4,5) \Phi(1,5,6)}{\Phi(1,3)\Phi(1,4)\Phi(1,5)}. 
\end{equation*}
The phase of the six-point invariant is therefore the sum of the four oriented triangle phases because each two-point invariant in the denominator is real and non-negative.

\subsubsection{Wilson loop} \label{subsec:wilson}
Under appropriate symmetry constraints, a noncontractible loop can carry quantized holonomy. Discretize a closed path as $\vb{k}_0\to\vb{k}_1\to\cdots\to\vb{k}_L\equiv\vb{k}_0$ and define the Wilson-loop matrix 
\begin{equation}
\mathcal{W}_{\mathrm{loop}} = M_{0,1}M_{1,2}M_{2,3} \cdots M_{L-1,L}
\end{equation}
where $M_{i,i+1}=\Psi(\vb{k}_i)^\dagger\Psi(\vb{k}_{i+1})$. At finite mesh spacing these overlaps are generally not unitary. The product converges to the unitary Wilson loop in the continuum limit; alternatively, one may replace each overlap by its unitary polar factor. 
The Wilson loop spectrum is related to the Wannier centers of occupied bands~\cite{Neupert_2018,Bradlyn2022}.
For the raw finite-mesh product, the gauge-invariant total Berry phase is
\begin{equation}
e^{-i \varphi_B} = \frac{\det \mathcal{W}_{\mathrm{loop}}}
{|\det \mathcal{W}_{\mathrm{loop}}|},
\end{equation}
whenever the determinant is nonzero. In the continuum limit,
\begin{equation}
 \varphi_B = i\int_{\vb{k}_0}^{\vb{k}_L} \trace \mathcal{A}= i\oint \trace \mathcal{A}  \qq{mod $2\pi$,}
\end{equation}
where the second equality emphasizes that the result is independent of the choice of base point.

Our aim is to evaluate this phase via projectors without choosing a smooth gauge. Although the determinant of the Wilson-loop matrix admits a triangulation analogous to Eq.~(\ref{barg_decomp}), it is not itself the trace of a single ordered product for $k>1$. 
To this end, define the ordered projector product
\begin{equation}
 P_{\mathrm{loop}} = P(\vb{k}_0)  P(\vb{k}_1) \cdots P(\vb{k}_L),
\end{equation}
whose restriction to the occupied subspace at the base point is precisely an element of the Wilson-loop matrix:
\begin{equation}
	(\mathcal{W}_{\mathrm{loop}})_{\alpha \beta} = \matrixel{u_\alpha(\vb{k}_0)}{P_{\mathrm{loop}}}{u_\beta(\vb{k}_0)} .
\end{equation}
A direct expansion of $\det\mathcal W_{\mathrm{loop}}$ could introduce projectors onto individual bands,
\begin{equation}
	d_\alpha (\vb{k})=u_\alpha(\vb{k}) u_\alpha(\vb{k})^\dagger, 
\end{equation}
so that $P(\vb{k}) = \sum_{\alpha=1}^k d_\alpha (\vb{k})$. 
Defining 
\begin{equation}
D_\alpha = d_\alpha (\vb{k}_0) P_{\mathrm{loop}} ,
\end{equation}  
one finds, for instance, 
\begin{equation}
	\det \mathcal{W}_{\mathrm{loop}} = \trace (D_1)  \qquad  k=1,
\end{equation}
\begin{equation}
	\det \mathcal{W}_{\mathrm{loop}} = \trace (D_1)\trace (D_2) -\trace (D_1 D_2) \qquad k=2. 
\end{equation}
The expressions become increasingly complicated as the number of occupied bands increases. 
This construction, however, is not applicable in the presence of degeneracies. For example, at time-reversal-invariant momenta the Kramers-degenerate subspace cannot be decomposed uniquely into individual band projectors $d_\alpha$. 
Fortunately, such a decomposition is unnecessary. 
As shown in Appendix~\ref{appendix:det_Wilson}, the determinant can be expressed solely through $\trace(P_{\mathrm{loop}}^m)=\trace (\mathcal{W}_{\mathrm{loop}}^m)$ for $m\leq k$. For example,
\begin{equation}
	\det \mathcal{W}_{\mathrm{loop}} = \trace P_{\mathrm{loop}} \qquad  k=1,
\end{equation}
\begin{equation}
	\det \mathcal{W}_{\mathrm{loop}} 
	= \frac{1}{2} \left[\tr(P_\text{loop})^2-\tr(P_\text{loop}^2) \right] \qquad k=2. 
\end{equation}
The general determinant and explicit low-rank cases are given in Appendix~\ref{appendix:det_Wilson}. The phase of this projector-only expression remains well defined in the presence of symmetry-protected degeneracies. 
This construction concerns the determinant of the Wilson-loop matrix; recovering its full eigenvalue spectrum requires all characteristic-polynomial coefficients by the Cayley–Hamilton theorem, which are likewise determined by the same power traces. 
Although the closeness between $\mathcal{W}_{\mathrm{loop}}$ and $P_{\mathrm{loop}}$ is well-known, this result provides an explicit projector-only formulation of the Wilson-loop determinant that remains applicable at internal band degeneracies.

\section{Differential forms} \label{sec:diff_form}
We derive the differential forms of quantum states by applying the exterior derivative operator $\dd$ on eigenfunctions. 
The full connection in the complete occupied–unoccupied frame is
\begin{align}
    \begin{split}
        \widetilde{\mathcal{A}} &= \mqty(\Psi^\dagger \\ \Psi_\perp^\dagger) \dd \mqty(\Psi & \Psi_\perp) = \mqty(\Psi^\dagger \dd \Psi & \Psi^\dagger \dd \Psi_\perp \\ \Psi_\perp^\dagger \dd \Psi & \Psi_\perp^\dagger \dd \Psi_\perp) \\
        &=  \mqty(\mathcal{A} & -\mathcal{B}^\dagger \\ \mathcal{B} &  \mathcal{A}_\perp) .
    \end{split}
\end{align}
Here $\mathcal A=\Psi^\dagger\dd\Psi$ and $\mathcal A_\perp=\Psi_\perp^\dagger\dd\Psi_\perp$ are skew-Hermitian intraband connections, while $\mathcal B=\Psi_\perp^\dagger\dd\Psi$ is the interband one-form. The block structure is the differential-form counterpart of the Cartan decomposition.

Completeness ($\Psi \Psi^\dagger + \Psi_\perp \Psi_\perp^\dagger = I_N$) and orthonormality ($\Psi^\dagger \Psi = I_k$ and $\Psi_\perp^\dagger \Psi_\perp = I_{N-k}$) give
\begin{align}
    \begin{split}
        \dd \Psi &= \dd \Psi \Psi^\dagger \Psi  \\
&= -(\Psi \dd \Psi^\dagger + \Psi_\perp \dd \Psi_\perp^\dagger + \dd \Psi_\perp  \Psi_\perp^\dagger )\Psi \\
&=-\Psi \dd \Psi^\dagger \Psi - \Psi_\perp \dd \Psi_\perp^\dagger \Psi \\
& = \Psi \mathcal{A} + \Psi_\perp \mathcal{B}, \label{dPsi}
    \end{split}
\end{align}
and
\begin{align}
    \begin{split}
        \dd \Psi_\perp &= \dd \Psi_\perp \Psi_\perp^\dagger \Psi_\perp  \\
&= -(\Psi \dd \Psi^\dagger + \Psi_\perp \dd \Psi_\perp^\dagger + \dd \Psi  \Psi^\dagger )\Psi_\perp \\
&=-\Psi \dd \Psi^\dagger \Psi_\perp - \Psi_\perp \dd \Psi_\perp^\dagger \Psi_\perp \\
& = - \Psi \mathcal{B}^\dagger + \Psi_\perp \mathcal{A}_\perp.
    \end{split}
\end{align}
Equivalently, in matrix form,
\begin{align}
    \begin{split}
        \dd \mqty(\Psi & \Psi_\perp) = \mqty(\Psi & \Psi_\perp) \mqty(\mathcal{A} & -\mathcal{B}^\dagger \\ \mathcal{B} &  \mathcal{A}_\perp)= \mqty(\Psi & \Psi_\perp) \widetilde{\mathcal{A}}, 
    \end{split}
\end{align}
and
\begin{align}
    \begin{split}
        \dd \mqty(\Psi^\dagger \\ \Psi_\perp^\dagger) = -\mqty(\mathcal{A} & -\mathcal{B}^\dagger \\ \mathcal{B} &  \mathcal{A}_\perp) \mqty(\Psi^\dagger \\ \Psi_\perp^\dagger) = - \widetilde{\mathcal{A}} \mqty(\Psi^\dagger \\ \Psi_\perp^\dagger) .
    \end{split}
\end{align}
Because $\dd^2=0$, the complete-basis connection satisfies the Maurer--Cartan equation:
\begin{align}
    \begin{split}
        \dd \widetilde{\mathcal{A}} &= \mqty(\dd \mathcal{A} & -\dd \mathcal{B}^\dagger \\ \dd \mathcal{B} &  \dd \mathcal{A}_\perp) \\
        &=\dd \mqty(\Psi^\dagger \\ \Psi_\perp^\dagger) \wedge \dd \mqty(\Psi & \Psi_\perp) \\
        & = - \widetilde{\mathcal{A}}\wedge \widetilde{\mathcal{A}} \\
        & = \mqty(-\mathcal{A}\wedge \mathcal{A} + \mathcal{B}^\dagger \wedge \mathcal{B}  & \mathcal{A}\wedge \mathcal{B}^\dagger + \mathcal{B}^\dagger \wedge \mathcal{A}_\perp \\ -\mathcal{B} \wedge \mathcal{A} - \mathcal{A}_\perp \wedge \mathcal{B}   & \mathcal{B} \wedge \mathcal{B}^\dagger - \mathcal{A}_\perp \wedge \mathcal{A}_\perp ).
    \end{split} \label{totalA}
\end{align}
Comparing blocks gives
\begin{align}
    \mathcal{F} & \equiv \dd \mathcal{A}+ \mathcal{A} \wedge \mathcal{A} =\mathcal{B}^\dagger\wedge \mathcal{B}, \label{F} \\
    \mathcal{F}_\perp & \equiv \dd \mathcal{A}_\perp+ \mathcal{A}_\perp \wedge \mathcal{A}_\perp =\mathcal{B}\wedge \mathcal{B}^\dagger , \\
    \dd \mathcal{B} & = -\mathcal{B} \wedge \mathcal{A} - \mathcal{A}_\perp \wedge \mathcal{B}.  \label{dB}
\end{align}
To systematically handle orientation on the manifold, we introduce the wedge product $\wedge$, which satisfies $\dd k^\mu \wedge \dd k^\nu = - \dd k^\nu \wedge \dd k^\mu$ upon exchanging any two scalar 1-forms. On matrix-valued 1-form, $(\mathcal{B} \wedge \mathcal{A})^\dagger = -\mathcal{A}^\dagger \wedge \mathcal{B}^\dagger = \mathcal{A} \wedge \mathcal{B}^\dagger$.

Equation~(\ref{F}) is the non-Abelian Berry curvature of the occupied bundle. With the covariant exterior derivative $\mathcal D=\dd+[\mathcal A,\cdot]$, Eqs.~(\ref{F}) and (\ref{dB}) imply the Bianchi identity $\mathcal D\mathcal F=0$. The matrix-valued form $\mathcal F$ need not be closed under $\dd$, but its invariant polynomials are closed.

Under $\Psi\mapsto\Psi U_k$ and $\Psi_\perp\mapsto\Psi_\perp U_{N-k}$, the blocks transform as $\mathcal A\mapsto U_k^\dagger\mathcal A U_k+U_k^\dagger\dd U_k$, $\mathcal A_\perp\mapsto U_{N-k}^\dagger\mathcal A_\perp U_{N-k}+U_{N-k}^\dagger\dd U_{N-k}$, and $\mathcal B\mapsto U_{N-k}^\dagger\mathcal B U_k$. The inhomogeneous term $\omega=U_k^\dagger\dd U_k$ is pure gauge and satisfies
\begin{equation}
    \dd \omega + \omega \wedge \omega=0. 
\end{equation}
The curvature transforms covariantly, $\mathcal F\mapsto U_k^\dagger\mathcal F U_k$. Consequently, $\trace(\mathcal F^n)$ is gauge invariant, and the Bianchi identity implies $\dd\trace(\mathcal F^n)=0$. These closed forms represent the Chern character of the occupied bundle~\cite{nakahara2018geometry,avron1989chern}.

The complete connection is flat, $\dd\widetilde{\mathcal A}+\widetilde{\mathcal A}\wedge\widetilde{\mathcal A}=0$, because it is a pure change of the full frame. The occupied and unoccupied curvature matrices have different dimensions and are not negatives of one another; rather, their trace curvatures cancel:
\begin{equation}
 \trace\mathcal F+\trace\mathcal F_\perp=0.
\end{equation}
This is analogous to the Newton’s third law in the sense that the geometric force on the (whole) occupied bands is of equal magnitude and opposite direction to that on the unoccupied bands.

On a patch supporting a smooth frame, $\trace(\mathcal A\wedge\mathcal A)=0$, and Stokes' theorem gives
\begin{equation}
    \varphi_B(C) =  i\int_S \trace \mathcal{F}=i\oint_{C=\partial S} \trace \mathcal{A},
\end{equation}
The equality holds modulo $2\pi$ at the level of the physical phase, arising from the ambiguity term $i \oint \trace \omega$. For a closed domain $S$, if no single smooth gauge exists on $S$, this obstruction allows a nonzero Chern number. On the other hand, when $S$ is not simply connected or noncontractible, flat connections can have nontrivial holonomy around noncontractible cycles. The BZ is a torus and therefore has such cycles, making Wilson loops natural probes of polarization and symmetry-protected topology.

\section{Topological invariants}\label{sec:topology}

\subsection{Chern number} \label{subsec:2form}
The canonical curvature two-form of the projector connection is
\begin{equation}
    \mathcal{F}_P = P \dd{P} \wedge \dd{P}, \label{F_P}
\end{equation}
where $\dd P = \partial_\mu P \dd k^\mu$, the differential of the projector map: $\dd P_{\vb{k}}:T_{\vb{k}}\mathcal M\to T_{P(\vb{k})}\mathrm{Gr}(k,N)$ for $\vb{k}\in \mathrm{BZ}$. 
Because $P\dd P P=0$, one equivalently has $\mathcal F_P=P\dd P\wedge\dd P P$. 
Writing $\dd P=[\mathcal X,P]$ in terms of the horizontal generator gives $\mathcal F_P=-P\mathcal X\wedge\mathcal X P$.
Substituting $\mathcal{X} = \Psi_\perp \mathcal{B} \Psi^\dagger -  \Psi \mathcal{B}^\dagger \Psi_\perp^\dagger$, we obtain 
\begin{equation}
    \mathcal{F}_P =  \Psi \mathcal{F} \Psi^\dagger.
\end{equation}
In contrast to the local curvature 2-form $\mathcal{F}$, which transforms covariantly as $\mathcal{F} \mapsto U_k^\dagger \mathcal{F} U_k$, the 2-form $\mathcal{F}_P$ constructed from the projector is gauge-invariant.

Taking the trace gives the Abelian Berry-curvature two-form
\begin{align}
\begin{split}
    F^{(2)} &= i \trace \mathcal{F} 
    = i \trace \left( \dd \mathcal{A} +\mathcal{A} \wedge \mathcal{A} \right) \\
        &= \frac{1}{2} F_{\mu \nu} \dd k^\mu \wedge \dd k^\nu,
\end{split}   
\end{align}
where $\mathcal{A}=\Psi^\dagger \dd \Psi = -i A_\mu \dd{k}^\mu $, and the gauge-field Berry connection is defined as
\begin{equation}
    (A_\mu)_{mn} = i u_m^\dagger \partial_\mu u_n= i \braket{m}{\partial_\mu n}   .
\end{equation}
Here, the imaginary factor $i$ is introduced to ensure Hermiticity. The corresponding Berry curvature tensor $F_{\mu \nu}$ is then given by 
\begin{equation}
    F_{\mu \nu} = \trace \left( \partial_\mu A_\nu -  \partial_\nu A_\mu -i \comm{A_\mu}{A_\nu} \right).
\end{equation}
Expressing the matrix components of the Berry curvature tensor explicitly yields
\begin{align}
\begin{split}
    F_{\mu \nu} &= i \sum_{m\in \mathrm{val}} \sum_{l\in \mathrm{cond}} 
    \braket{\partial_\mu m}{l}  \braket{l}{\partial_\nu m} -  (\mu \leftrightarrow \nu) \\
     &= i \sum_{m\in \mathrm{val}} \sum_{l\in \mathrm{cond}} (B_{\mu}^*)^{lm} B_{\nu}^{lm} -  (\mu \leftrightarrow \nu) \\
    &= - 2 \Im \trace \left( B_{\mu}^{\dagger} B_{\nu} \right) , \label{F_berry}
\end{split}   
\end{align}
where the interband Berry connection between the occupied and unoccupied bands is defined as 
\begin{equation}
    (B_\mu)_{l m} = \braket{l}{\partial_\mu m} , \qq{where} u_m\in \Psi~ \qq*{and}u_l\in \Psi_\perp .
\end{equation}
The final equality in Eq.~(\ref{F_berry}) demonstrates that the matrix $B$ in the tangent vector formulation of Eq.~(\ref{Tt}) corresponds precisely to the interband Berry connection.

The scalar form $\trace\mathcal F_P$ is closed but need not be globally exact. More generally, in $2n$ dimensions the degree-$2n$ component of the Chern character is
\begin{equation}
    F^{(2n)} = \trace[(i\mathcal{F}_P)^{n}]= \trace[(i\mathcal{F})^{n}],
\end{equation}
and the corresponding $n$th Chern number is
\begin{align}
\begin{split}
    \mathrm{Ch}_n &= \frac{1}{n!} \left(\frac{1}{2\pi}\right)^n \int_{\mathrm{BZ}} F^{(2n)}  \\
    &= \frac{1}{n!} \left(\frac{i}{2\pi}\right)^n  \epsilon^{\alpha_1 \alpha_2 \cdots \alpha_{2n}} \\
    & ~~~~~~~~~  \times  \int_{\mathrm{BZ}}  \trace \left( B_{\alpha_1}^\dagger B_{\alpha_{2}} \cdots B_{\alpha_{2n-1}}^\dagger B_{\alpha_{2n}} \right)  \dd[2n]{k} ,
\end{split}   
\end{align}
where the matrix products are antisymmetrized by the Levi-Civita tensor. These expressions remain meaningful without choosing a globally smooth occupied frame.

\subsection{Chiral symmetry} \label{subsec:winding}
Consider a gapped system with a momentum-independent chiral operator $\Gamma~(=\Gamma^\dagger)$, where $\Gamma^2=I_{2k}$ and
\begin{equation}
    \Gamma H(\vb{k}) \Gamma^\dagger = -H(\vb{k}).
\end{equation}
We note that the numbers of the occupied and unoccupied bands are both $k$. 
In the chiral-paired frame, the chiral operator $\Gamma$ interchanges the occupied and the unoccupied states: $\Gamma \Psi = \Psi_\perp$ and $\Gamma \Psi_\perp = \Psi$. 
We define two subspaces
\begin{equation}
\Psi_{\Gamma_+} = \frac{1}{\sqrt{2}}(\Psi_\perp + \Psi) \quad \text{and} \quad \Psi_{\Gamma_-} = \frac{1}{\sqrt{2}}(\Psi_\perp - \Psi),
\end{equation}
which correspond to the eigenstates of $\Gamma$ with eigenvalues $+1$ and $-1$, respectively. In a practical manner, an arbitrary state can be projected into a chirality subspace via the projectors $\Pi_\pm=(I_{2k}\pm\Gamma)/2$. 	

Chiral symmetry implies $\mathcal A_\perp=\mathcal A$ and $\mathcal B=-\mathcal B^\dagger$. Inserting the identity $\Gamma^\dagger \Gamma = I_{2k}$ into the interband connection $(B_\mu)^{m n}=\braket{u^{\text{c}}_m}{\partial_\mu u^{\text{v}}_n} $ with $u^{\text{c}}_m \in \Psi_\perp$ and $u^{\text{v}}_n \in \Psi$, we have $(B_\mu)^{m n}= \braket{\Gamma u^{\text{c}}_m}{\partial_\mu (\Gamma  u^{\text{v}}_n)} = \braket{u^{\text{v}}_m}{\partial_\mu  u^{\text{c}}_n}=-(B_\mu^\dagger)^{mn}$. 
Similarly, the intraband connections satisfy $A_\mu = A_{\perp \mu}$. 

The chiral-paired condition restricts the gauge freedom to simultaneous transformations of the occupied and unoccupied frames,
\begin{equation}
    \Psi\mapsto\Psi U_k,
    \qquad
    \Psi_\perp\mapsto\Psi_\perp U_k.
\end{equation}
Consequently, $\Psi_{\Gamma_{\pm}} \mapsto \Psi_{\Gamma_{\pm}} U_k$ and $B_\mu \mapsto U_k^\dagger B_\mu U_k $, without an inhomogeneous derivative term. As a result, the generator, in terms of differential form, $\mathcal{X} = \Psi_\perp \mathcal{B} \Psi^\dagger -  \Psi \mathcal{B}^\dagger \Psi_\perp^\dagger$ is invariant under both chiral and gauge transformations. 
Restricting \(\mathcal X\) to the negative-chirality subspace defines the gauge-invariant one-form 
\begin{equation}
\mathcal A_\Gamma
=-2\Pi_-\mathcal X\Pi_-
=\Psi_{\Gamma_-}(2\mathcal B)\Psi_{\Gamma_-}^\dagger.
\label{AG1}
\end{equation}

This one-form is actually a differential object associated with the chiral transition map. Introduce the chiral transition operator $q_\Gamma$,
\begin{equation}
	q_\Gamma: =\Psi_{\Gamma_+} \Psi_{\Gamma_-}^\dagger =-2\Pi_+ P \Pi_- ,
\end{equation}
that maps the negative-chirality subspace onto the positive-chirality subspace. It satisfies $q_\Gamma^\dagger q_\Gamma=\Pi_-$ and $q_\Gamma q_\Gamma^\dagger=\Pi_+$ and preserves norms on the negative-chirality subspace (a partial isometry). 
The connection on $q_\Gamma$ turns out to be $\mathcal{A}_\Gamma$: 
\begin{equation}
\mathcal A_\Gamma
=q_\Gamma^\dagger\dd q_\Gamma
=\Psi_{\Gamma_-}
 \left(\mathcal A_{\Gamma_+}-\mathcal A_{\Gamma_-}\right)
 \Psi_{\Gamma_-}^\dagger ,
\label{AG2}
\end{equation}
where the connections in the subspace bases are $\mathcal{A}_{\Gamma_\pm} = \Psi_{{\Gamma_\pm}}^\dagger \dd  \Psi_{{\Gamma_\pm}} = \frac{1}{2} \left[ \mathcal{A} + \mathcal{A}_\perp \pm  \left(\mathcal{B} - \mathcal{B}^\dagger \right) \right]= \mathcal{A} \pm \mathcal{B}$. 
In the above derivation, the cross-coupling term $\Psi_{{\Gamma_+}}^\dagger \dd \Psi_{{\Gamma_-}}$ in Eq.~(\ref{AG2}) vanishes identically due to chiral symmetry:
\begin{equation}
\Psi_{{\Gamma_+}}^\dagger \dd \Psi_{{\Gamma_-}} = (\Gamma \Psi_{{\Gamma_+}} )^\dagger \dd (\Gamma \Psi_{{\Gamma_-}} ) = -\Psi_{{\Gamma_+}}^\dagger \dd \Psi_{{\Gamma_-}}  = 0.
\end{equation}
Thus $\dd\Psi_{\Gamma_-}=\Psi_{\Gamma_-}\mathcal A_{\Gamma_-}$ within the negative-chirality subspace.

It is straightforward to show that the connection $\mathcal{A}_\Gamma$ is flat, satisfying the Maurer-Cartan equation:
\begin{equation}
   \dd \mathcal{A}_\Gamma + \mathcal{A}_\Gamma \wedge \mathcal{A}_\Gamma =0. \label{AGamma}
\end{equation}
This flatness condition implies that $\dd  \trace (\mathcal{A}_\Gamma^{2n+1}) = 0$. 
Therefore, $\trace(\mathcal A_\Gamma^{2n+1})$ is closed and defines a de Rham cohomology class. 
In odd dimensions, the topological invariant protected by chiral symmetry is the winding number, given by:
\begin{equation}
    \nu_{2n+1} = \frac{(-1)^n n!}{(2n+1)!}\left(\frac{i}{2\pi} \right)^{n+1}\int_{\mathrm{BZ}} \trace \left(\mathcal{A}_\Gamma^{2n+1} \right),\label{winding_AG}
\end{equation}
with the normalization chosen to give an integer for a smooth map on a closed $(2n+1)$-dimensional BZ. Appendix~\ref{appendix:expression_winding} proves equivalence with the conventional winding number expression~\cite{ryu2010}.

\subsection{Time-reversal symmetry}
We now consider spinful time-reversal (TR) symmetry, for which the occupied rank $k$ and total dimension $N$ are even. Let $\Theta=U_\Theta K$ be antiunitary, with $\Theta^2=-1$ and $U_\Theta=-U_\Theta^T$. Its action on an occupied basis is
\begin{equation}
    \Theta \Psi(\vb{k}) = \Psi(-\vb{k}) w(\vb{k}), \label{TRu}
\end{equation}
where $w_{nm}(\vb{k})=\matrixel{u_n(-\vb{k})}{\Theta}{u_m(\vb{k})}$ is the unitary sewing matrix. The condition $\Theta^2=-1$ implies $w(-\vb{k})=-w^T(\vb{k})$~\cite{fu2006,bernevig2013}. 
Under a general local basis change $\Psi(\vb{k})\mapsto\Psi(\vb{k})U(\vb{k})$, the sewing matrix transforms as
\begin{equation}
 w(\vb{k})\mapsto U(-\vb{k})^\dagger w(\vb{k})U(\vb{k})^*.
\end{equation}
If one restricts to transformations that preserve a chosen sewing matrix, the frames at opposite momenta obey
\begin{align}
    U(-\vb{k})=w(\vb{k})U^*(\vb{k})w^\dagger(\vb{k}).
\end{align}
The obstruction to choosing such a smooth TR-compatible basis globally is measured by the $\mathbb Z_2$ index.

The following transition matrix is invariant under the local gauge transformation:
\begin{equation}
    t_\Theta(\vb{k}) =\left[\Theta\Psi(\vb{k}) \right]^* \Psi(\vb{k})^\dagger = \Psi(-\vb{k})^* w(\vb{k})^* \Psi(\vb{k})^\dagger,
\end{equation}
and it satisfies
\begin{equation}
    t_\Theta(\vb{k})^\dagger  t_\Theta(\vb{k})= P(\vb{k}) 
      \qq{and} 
    t_\Theta(\vb{k}) t_\Theta(\vb{k})^\dagger = P(-\vb{k})^*.
\end{equation}
It can be generated via the projector:
\begin{equation}
    t_\Theta(\vb{k}) =K \Theta P(\vb{k}) = \left[ \Theta\Psi(\vb{k}) \right]^*   \Psi(\vb{k})^\dagger .
\end{equation}
Its logarithmic derivative is the gauge-invariant one-form
\begin{align}
    \begin{split}
        \mathcal{A}_{\Theta}(\vb{k}) &:=t_\Theta(\vb{k})^\dagger \dd t_\Theta(\vb{k}) \\
                &= \Psi(\vb{k}) \mathcal{B}(\vb{k})^\dagger \Psi_\perp(\vb{k})^\dagger,
    \end{split}
\end{align}
where Eq.~(\ref{dPsi}) has been used. In above, the following identity is considered~\cite{bernevig2013}:
\begin{equation}
\mathcal{A}(-\vb{k}) =w(\vb{k}) \left[\mathcal{A}(\vb{k})^* - \Omega(\vb{k}) \right] w(\vb{k})^\dagger,
\end{equation}
where $\Omega (\vb{k}) = w(\vb{k})^\dagger \dd w(\vb{k}) $ is the connection one-form of the sewing matrix.

We then introduce a projector combining the occupied subspaces at $\vb{k}$ and $-\vb{k}$:
\begin{align}
    \begin{split}
        P_\Theta(\vb{k}) = \phi(\vb{k}) \phi(\vb{k})^\dagger =\frac{1}{2} \mqty( P(\vb{k}) & t_\Theta(\vb{k})^\dagger \\ t_\Theta(\vb{k}) & P(-\vb{k})^* )
    \end{split}
\end{align}
where 
\begin{equation}
	\phi(\vb{k}) =  \frac{1}{\sqrt{2}}\mqty( \Psi(\vb{k}) \\  \left[ \Theta\Psi(\vb{k}) \right]^* ).
\end{equation}
By construction, $P_\Theta$ is invariant under the local gauge transformation, and it satisfies $P_\Theta^2=P_\Theta$. 
It therefore permits direct application of the projector formalism developed in Sec.~\ref{subsec:2form} to the paired TR subspaces.
Analogous to Eq.~(\ref{F_P}), we define the associated two-form as
\begin{align}
    \begin{split}
        \trace  \mathcal{F}_\Theta &=  \trace  \left( P_\Theta \dd{P_\Theta} \wedge \dd{P_\Theta} \right) \\ 
        & = \frac{1}{2} \trace \left[ \mathcal{F}(\vb{k}) -  \mathcal{F}(-\vb{k})\right] \\
        & =  \trace \mathcal{F}(\vb{k}),
    \end{split}
\end{align}
where the last equality follows from the time-reversal constraint: $ \mathcal{F}(\vb{k}) = -\left[ w(\vb{k})^\dagger  \mathcal{F}(-\vb{k})w(\vb{k})\right]^T$. 
Thus the local trace curvature of the paired construction reduces to that of the occupied-band projector on one member of each TR pair.

Although the first Chern number over the full Brillouin zone vanishes, a half Brillouin zone (HBZ) with a TR-compatible boundary gauge can carry the Fu--Kane invariant
\begin{equation}
 	\nu_{\text{FK}} = \frac{i}{2\pi} \left( \oint_{\partial \text{HBZ}} \tr \mathcal{A}  - \int_{\text{HBZ}} \tr \mathcal{F} \right) \quad \text{mod~2},
 \end{equation}
which measures the obstruction modulo two~\cite{fu2006,Moore2007}. The boundary term can be evaluated from the determinant of the Wilson loop using Sec.~\ref{subsec:wilson}. Recent work further shows that the winding of the full Wilson spectrum imposes a lower bound on the integrated quantum metric, including in TR-protected $\mathbb Z_2$ phases~\cite{Yu2025Wilson}.

\section{Conclusion}
We have formulated the geometry of an isolated rank-$k$ quantum subspace directly in terms of its spectral projector, viewing a band structure as a map into $\mathrm{Gr}(k,N)$. This choice is more than a change of notation: for a smooth gapped Hamiltonian, the projector is globally smooth and periodic even when topology obstructs a globally smooth periodic or symmetry-compatible Bloch frame. The formulation therefore relocates topology from gauge singularities of wave functions to global properties of the projector map. Its central object is the differential $\dd P$. Physically, $\dd P$ removes internal gauge redundancy and retains the interband connection. Geometrically, $\dd P$ maps parameter-space variations to tangent vectors of the Grassmannian. For topology, products and wedge products of $\dd P$ give efficient gauge-invariant representatives of characteristic classes. 

We used this tangent-vector picture to connect two finitely separated projectors along a shortest path in the ambient Grassmannian. The singular values of the horizontal generator block are the principal angles between the endpoint subspaces, giving $l=(2\sum_\alpha\theta_\alpha^2)^{1/2}$ in our metric convention. Separating positive overlap amplitudes from unitary polar factors gives a piecewise-geodesic interpretation of Bargmann phases. For a degenerate multiband subspace, the determinant of the Wilson loop is expressed entirely through traces of powers of an ordered projector product, without selecting individual bands. 

The differential-form calculus then organizes global topology. Even-dimensional Chern characters follow from powers of $P\,\dd P\wedge\dd P$, chiral symmetry produces a flat Maurer--Cartan form whose odd traces yield winding numbers, and time-reversal-related spaces can be assembled into an enlarged projector. The determinant of the boundary Wilson loop supplies the Abelian holonomy entering the Fu--Kane $\mathbb Z_2$ invariant. Metric distance, curvature, holonomy, and symmetry-protected topology thus appear as complementary aspects of a single projector map.

Recent measurements of the QGT and quantum-metric-driven nonlinear transport demonstrate that quantum geometry is becoming an experimentally resolved property of solids. Natural extensions are to relate nonlinear responses to higher-order Grassmannian geometry and to develop symmetry-constrained projector geometry for real and quaternionic band structures.

\section{Acknowledgments}
This work was supported by the National Science and Technology Council (NSTC) under Grant No.~115-2112-M-110-023. The author also acknowledges support from the National Center for Theoretical Sciences (NCTS).

\appendix
\section{Elements of the generator} \label{appendix:generator}
In this Appendix we derive the horizontal generator directly from the Hamiltonian. Let $u_i$ belong to the selected band subspace and $u_a$ to its complement. Differentiating $H u_i=\varepsilon_i u_i$ with respect to a parameter $t$ and projecting onto $\bra{u_a}$ gives
\begin{equation}
    \langle u_a|\dot u_i\rangle
    = \frac{\langle u_a|\dot H|u_i\rangle}{\varepsilon_i-\varepsilon_a}
    =: (X^{(0)})_{ai}. \label{Xgen}
\end{equation}
The spectral gap ensures that the denominator is nonzero. Degeneracies within either subspace cause no ambiguity because only matrix elements across the gap enter the projector derivative. Define $X^{(0)}$ by Eq.~\eqref{Xgen}, its skew-Hermitian conjugate block $(X^{(0)})_{ia}=-(X^{(0)})_{ai}^*$, and vanishing intraband blocks. In the complete eigenbasis $\widetilde\Psi=(u_1,\ldots,u_N)$, the corresponding ambient-space operator is
\begin{equation}
    X=\widetilde\Psi X^{(0)}\widetilde\Psi^\dagger,
\end{equation}
and direct differentiation yields
\begin{equation}
    \dot P=\widetilde\Psi[X^{(0)},P^{(0)}]\widetilde\Psi^\dagger=[X,P].
\end{equation}
Thus $X$ is precisely the unique horizontal generator introduced in Eq.~\eqref{Tmu0}. In the block convention of Eq.~\eqref{X0}, $(X^{(0)})_{ai}=B_{ai}$ and $(X^{(0)})_{ia}=-(B^\dagger)_{ia}$. Any intraband terms arising from a particular choice of eigenvector phases or frames commute with $P$ and therefore do not contribute to $\dot P$.

\section{Wilson loop in terms of projectors} \label{appendix:det_Wilson}
This Appendix complements Sec.~\ref{subsec:wilson}. We derive explicit expressions for the determinant of the Wilson-loop matrix in terms of the projectors onto the occupied bands. The derivation is based on the Cayley--Hamilton theorem, which allows the determinant of a matrix to be expressed entirely in terms of the traces of its powers. 

The Cayley--Hamilton theorem states that every square $n\times n$ matrix $A$ satisfies its own characteristic equation:
\begin{equation}
	p(A) = A^n - c_{1}A^{n-1} + \cdots + (-1)^n c_{n} I_n = 0, 
\end{equation}
where $p(\lambda) = \lambda^n - c_{1}\lambda^{n-1} + \cdots +  (-1)^n c_n$ is the characteristic polynomial of $A$.  
The coefficients $c_i$ ($c_0=1$ by definition) are given by the elementary symmetric polynomials of the eigenvalues $\{\lambda_i\}$ of $A$:
\begin{align}
    \begin{split}
        c_{1} &= \lambda_1 + \lambda_2 + \cdots + \lambda_n = \trace(A), \\
        c_{2} &= \lambda_1 \lambda_2 + \lambda_1 \lambda_3  + \cdots + \lambda_{n-1} \lambda_n, \\
        & ~~~~~~~~~~ ~~~~~~ \vdots \\
        c_{n} &=  \lambda_1 \lambda_2 \cdots \lambda_n = \det A.
    \end{split}
\end{align}
The coefficients can be constructed recursively from the power sums
\begin{equation}
 p_j = \tr (A^j) 
\end{equation}
through the Newton identities:
\begin{equation}
	j c_{j} = \sum_{i=1}^j(-1)^{i-1} c_{j-i} p_i   ~~~~~ (1\le j \le n).
\end{equation}
Thus, all coefficients, and in particular \(c_n=\det A\), can be determined recursively from the traces of powers of \(A\). 
Similar techniques have also been employed in the construction of eigenprojectors from a given Hamiltonian \cite{graf2021,huang2025gaussbonnet}. In the present problem, the Cayley--Hamilton theorem allows us to express the determinant of the $k \times k$ matrix $\mathcal{W}_{\mathrm{loop}}$ in terms of traces of its powers. Furthermore, because $\trace (\mathcal{W}_{\mathrm{loop}}^m) = \trace (P_\text{loop}^{m})$, the Berry phase associated with the Wilson loop can be expressed solely in terms of the projectors. Therefore, for arbitrary $k$, the determinant of $\mathcal{W}_{\mathrm{loop}}$ is  
\begin{widetext}
\begin{equation}
\det \mathcal{W}_{\mathrm{loop}} =
\frac{1}{k!}
\mdet{
\trace (P_\text{loop})      & k-1              & 0                & \cdots & 0 \\
\trace (P_\text{loop}^{2})  & \trace (P_\text{loop}) & k-2            & \cdots & 0 \\
\vdots                  & \vdots           & \ddots           & \ddots & \vdots \\
\trace (P_\text{loop}^{k-1})& \trace (P_\text{loop}^{k-2}) & \cdots & \trace (P_\text{loop}) & 1 \\
\trace (P_\text{loop}^{k} ) & \trace (P_\text{loop}^{k-1}) & \cdots & \trace (P_\text{loop}^{2}) & \trace (P_\text{loop} )
}.
\end{equation}
For completeness, we list the explicit expressions for \(k=1\), \(2\), \(3\), and \(4\), where \(k\) denotes the number of occupied bands.
\subsubsection*{$k=1$.}
\begin{align}
    \det \mathcal{W}_{\mathrm{loop}} = \trace (D_1) = \trace P_{\mathrm{loop}} 
\end{align}

\subsubsection*{$k=2$.}
\begin{align}
\begin{split}
\det \mathcal{W}_{\mathrm{loop}} 
&= \trace (D_1)\trace (D_2) -\trace (D_1 D_2) \\
&= \frac{1}{2} \left[\tr(P_\text{loop})^2-\tr(P_\text{loop}^2) \right], \\
\end{split}
\end{align}

\subsubsection*{$k=3$.}
\begin{align}
\begin{split}
\det \mathcal{W}_{\mathrm{loop}} &= \trace (D_1)\trace (D_2) \trace (D_3)  -\trace (D_1) \trace (D_2 D_3)\\
  & ~~~~  -\trace (D_2) \trace (D_1 D_3) -\trace (D_3) \trace (D_1 D_2) \\
  & ~~~~ +\trace (D_1 D_2 D_3) +\trace (D_1 D_3 D_2) \\
 & = \frac{1}{6} \left[ \tr(P_\text{loop})^3 -3 \tr(P_\text{loop})\tr(P_\text{loop}^2) +2 \tr(P_\text{loop}^3) \right],
\end{split}
\end{align}

\subsubsection*{$k=4$.}
\begin{align}
\begin{split}
\det \mathcal{W}_{\mathrm{loop}} &= \tr(D_1) \tr(D_2) \tr(D_3) \tr(D_4) \\
&- \tr(D_1) \tr(D_2) \tr(D_3 D_4) - \tr(D_1) \tr(D_3) \tr(D_2 D_4) 
- \tr(D_1) \tr(D_4) \tr(D_2 D_3) \\
&- \tr(D_2) \tr(D_3) \tr(D_1 D_4) - \tr(D_2) \tr(D_4) \tr(D_1 D_3) - \tr(D_3) \tr(D_4) \tr(D_1 D_2) \\
& +  \tr(D_1 D_2) \tr(D_3 D_4)  +  \tr(D_1 D_3) \tr(D_2 D_4) +  \tr(D_1 D_4) \tr(D_2 D_3) \\
 & +  \tr(D_1) \left[ \tr(D_2  D_3 D_4) +  \tr(D_2  D_4 D_3) \right]  +  \tr(D_2) \left[ \tr(D_1  D_3 D_4) +  \tr(D_1  D_4 D_3) \right] \\
 & +  \tr(D_3) \left[ \tr(D_1  D_2 D_4) +  \tr(D_1  D_4 D_2) \right]  +  \tr(D_4) \left[ \tr(D_1  D_2 D_3) +  \tr(D_1  D_3 D_2) \right] \\
 & -  \tr(D_1 D_2  D_3 D_4)  -  \tr(D_1 D_2  D_4 D_3)  -  \tr(D_1 D_3  D_2 D_4) \\
 & -  \tr(D_1 D_3  D_4 D_2)  -  \tr(D_1 D_4  D_2 D_3) -  \tr(D_1 D_4  D_3 D_2) \\
&=\frac{1}{24}\left[
\tr (P_\text{loop})^4
-6\,\tr (P_\text{loop}^2)(\tr P_\text{loop})^2
+3 \tr(P_\text{loop}^2) ^2
+8\,\tr (P_\text{loop}^3)\,\tr (P_\text{loop})
-6\,\tr (P_\text{loop}^4)
\right].
\end{split}
\end{align}

\end{widetext}

\section{Conventional expression for the winding number} \label{appendix:expression_winding}
In this Appendix we briefly review the conventional expression for the winding number of a band insulator with chiral symmetry, $\{\Gamma, H \}=0$, defined over the BZ. The derivation follows the standard treatment in Ref.~\cite{ryu2010}. 
In the basis of $\Gamma=\mqty( I_k & 0 \\ 0 & -I_k)$, the Hamiltonian can be written as
\begin{equation}
    H = \mqty( 0 & \mathfrak{D} \\ \mathfrak{D}^\dagger & 0).
\end{equation}
Applying SVD, $\mathfrak{D} = \chi \Sigma \phi^\dagger $, where $\chi,~\phi\in \mathrm{U}(k)$. The columns of $\chi$ and $\phi$, denoted by $\chi_m$ and $\phi_m$, make up the positive- and negative-energy (conduction and valence) eigenstates: $\ket{u_m^{\text{c}}} =\frac{1}{\sqrt{2}}\mqty( \chi_m \\ \phi_m)$ and $\ket{u_m^{\text{v}}} =\frac{1}{\sqrt{2}} \mqty( \chi_m \\ -\phi_m)$ with eigenvalues $\Sigma_{mm}$ and $-\Sigma_{mm}$, respectively. Note that we operate in the chiral-paired frame: $\Gamma \ket{u_m^{\text{v}}}= \ket{u_m^{\text{c}}}$ and $\Gamma \ket{u_m^{\text{c}}} = \ket{u_m^{\text{v}}}$.

The difference of eigenprojectors (flattened Hamiltonian) is 
\begin{equation}
    \mathfrak{H}=\sum_m \left( \dyad{u_m^{\text{c}}} - \dyad{u_m^{\text{v}}}\right) = \mqty(0 & q \\ q^\dagger & 0), \label{QmP}
\end{equation}
where 
\begin{equation}
    q= \sum_m  \chi_m \phi_m^\dagger= \chi \phi^\dagger \in \mathrm{U}(k). \label{small_q}
\end{equation}
A SVD need not provide globally smooth \(\chi\) and \(\phi\). The unitary $q= \mathfrak{D} (\mathfrak{D}^\dagger \mathfrak{D} )^{-1/2} $, however, is globally defined because \(\mathfrak D\) is nonsingular. 
Thus, the quantum system is a mapping: $\mathrm{BZ} \to \mathrm{U}(k)$, whose homotopy class is characterized by the winding number. 
A connection 1-form is defined as
\begin{equation}
    \omega = q^\dagger \dd q = \phi \left( \chi^\dagger \dd \chi - \phi^\dagger \dd \phi \right) \phi^\dagger.
\end{equation}
The winding number in terms of $\omega$ is
\begin{equation}
    \nu_{2n+1} = \frac{(-1)^n n!}{(2n+1)!}\left(\frac{i}{2\pi} \right)^{n+1}\int_{\mathrm{BZ}} \trace \left(\omega^{2n+1} \right). \label{winding_w}
\end{equation}

We next establish the equivalence between this expression and the projector representation in Eq.~\eqref{winding_AG} of Sec.~\ref{subsec:winding}. 
The chiral eigenstates associated with the paired conduction and valence states are
\begin{equation}
    \ket{u_m^{(+)}} = \frac{1}{\sqrt{2}}\left( \ket{u_m^{\text{c}}} + \ket{u_m^{\text{v}}}\right) = \mqty(\chi_m \\ 0)  
\end{equation}
and
\begin{equation}
\ket{u_m^{(-)}} = \frac{1}{\sqrt{2}}\left( \ket{u_m^{\text{c}}} - \ket{u_m^{\text{v}}}\right) = \mqty(0 \\ \phi_m) .
\end{equation}
Consequently,
\begin{equation}
    q_\Gamma
    =\Psi_{\Gamma_+}\Psi_{\Gamma_-}^\dagger
    =\mqty(0&q\\0&0).
\end{equation}
It follows that
\begin{equation}
    \mathcal A_\Gamma
    =q_\Gamma^\dagger\dd q_\Gamma
    =\mqty(0&0\\0&q^\dagger\dd q)
    =\mqty(0&0\\0&\omega).
\end{equation}
Therefore,
\begin{equation}
    \trace(\mathcal A_\Gamma^{2n+1})
    =\trace(\omega^{2n+1}),
\end{equation}
which proves the equivalence of Eqs.~\eqref{winding_AG} and
\eqref{winding_w}.

This identification is independent of the diagonal representation.
Indeed, let $\Phi_\pm$ be momentum-independent orthonormal frames of
the fixed chiral subspaces, satisfying
\begin{equation}
	\Gamma\Phi_\pm=\pm\Phi_\pm,\qquad \Phi_\pm^\dagger\Phi_\pm=I_k,\qquad \Phi_\pm\Phi_\pm^\dagger=\Pi_\pm,
\end{equation} 
and define
\begin{equation}
    Q_\Gamma=\Phi_+^\dagger q_\Gamma\Phi_-.
\end{equation}
The partial-isometry identities
$q_\Gamma^\dagger q_\Gamma=\Pi_-$ and
$q_\Gamma q_\Gamma^\dagger=\Pi_+$ imply that
$Q_\Gamma\in\mathrm{U}(k)$. Moreover,
\begin{equation}
    \mathcal A_\Gamma
    =\Phi_-\left(Q_\Gamma^\dagger\dd Q_\Gamma\right)\Phi_-^\dagger.
\end{equation}
In the diagonal chiral basis, $Q_\Gamma=q$ exactly; arbitrary
momentum-independent changes of the two fixed chiral bases only
conjugate $Q_\Gamma^\dagger\dd Q_\Gamma$ by a constant unitary matrix
and therefore leave the winding number unchanged.

\bibliography{grass.bib}

@article{batzies_2015,
  author        = {Batzies, E. and Hüper, K. and Machado, L. and Leite, F. Silva},
  title         = {Geometric mean and geodesic regression on {Grassmannians}},
  journal       = {Linear Algebra Appl.},
  volume        = {466},
  pages         = {83--101},
  year          = {2015},
  doi           = {10.1016/j.laa.2014.10.003},
  url           = {https://doi.org/10.1016/j.laa.2014.10.003},
}

@book{lee2018introduction,
  author        = {Lee, John M.},
  title         = {Introduction to {Riemannian} Manifolds},
  edition       = {2nd},
  series        = {Graduate Texts in Mathematics},
  volume        = {176},
  publisher     = {Springer},
  address       = {Cham, Switzerland},
  year          = {2018},
  isbn          = {978-3-319-91755-9},
  doi           = {10.1007/978-3-319-91755-9},
  url           = {https://doi.org/10.1007/978-3-319-91755-9},
}

@article{Chen_2025,
  author        = {Chen, Wei},
  title         = {Quantum geometrical properties of topological materials},
  journal       = {J. Phys.: Condens. Matter},
  volume        = {37},
  number        = {2},
  pages         = {025605},
  year          = {2024},
  doi           = {10.1088/1361-648X/ad8619},
  url           = {https://doi.org/10.1088/1361-648X/ad8619},
}

@article{Resta2000,
  author        = {Resta, Raffaele},
  title         = {Manifestations of {Berry}'s phase in molecules and condensed matter},
  journal       = {J. Phys.: Condens. Matter},
  volume        = {12},
  number        = {9},
  pages         = {R107--R143},
  year          = {2000},
  doi           = {10.1088/0953-8984/12/9/201},
  url           = {https://doi.org/10.1088/0953-8984/12/9/201},
}

@incollection{berry1989quantum,
  author        = {Berry, M. V.},
  title         = {The quantum phase, five years after},
  booktitle     = {Geometric Phases in Physics},
  editor        = {Shapere, Alfred and Wilczek, Frank},
  series        = {Advanced Series in Mathematical Physics},
  volume        = {5},
  publisher     = {World Scientific},
  address       = {Singapore},
  pages         = {7--28},
  year          = {1989},
  isbn          = {978-9971-5-0599-8},
  doi           = {10.1142/9789812798381_0001},
  url           = {https://doi.org/10.1142/9789812798381_0001},
}

@article{wang2021,
  author        = {Wang, Jie and Cano, Jennifer and Millis, Andrew J. and Liu, Zhao and Yang, Bo},
  title         = {Exact Landau Level Description of Geometry and Interaction in a Flatband},
  journal       = {Phys. Rev. Lett.},
  volume        = {127},
  number        = {24},
  pages         = {246403},
  year          = {2021},
  doi           = {10.1103/PhysRevLett.127.246403},
  url           = {https://doi.org/10.1103/PhysRevLett.127.246403},
}

@article{Ozawa2021,
  author        = {Ozawa, Tomoki and Mera, Bruno},
  title         = {Relations between topology and the quantum metric for Chern insulators},
  journal       = {Phys. Rev. B},
  volume        = {104},
  number        = {4},
  pages         = {045103},
  year          = {2021},
  doi           = {10.1103/PhysRevB.104.045103},
  url           = {https://doi.org/10.1103/PhysRevB.104.045103},
}

@article{ryu2010,
  author        = {Ryu, Shinsei and Schnyder, Andreas P and Furusaki, Akira and Ludwig, Andreas WW},
  title         = {Topological insulators and superconductors: tenfold way and dimensional hierarchy},
  journal       = {New J. Phys.},
  volume        = {12},
  number        = {6},
  pages         = {065010},
  year          = {2010},
  doi           = {10.1088/1367-2630/12/6/065010},
  url           = {https://doi.org/10.1088/1367-2630/12/6/065010},
}

@article{fu2006,
  author        = {Fu, Liang and Kane, C. L.},
  title         = {Time reversal polarization and a ${Z}_{2}$ adiabatic spin pump},
  journal       = {Phys. Rev. B},
  volume        = {74},
  number        = {19},
  pages         = {195312},
  year          = {2006},
  doi           = {10.1103/PhysRevB.74.195312},
  url           = {https://doi.org/10.1103/PhysRevB.74.195312},
}

@article{panati2007triviality,
  author        = {Panati, Gianluca},
  title         = {Triviality of {Bloch} and {Bloch--Dirac} Bundles},
  journal       = {Ann. Henri Poincar\'e},
  volume        = {8},
  number        = {5},
  pages         = {995--1011},
  year          = {2007},
  doi           = {10.1007/s00023-007-0326-8},
  url           = {https://doi.org/10.1007/s00023-007-0326-8},
}

@article{Avdoshkin2023,
  author        = {Avdoshkin, Alexander and Popov, Fedor K.},
  title         = {Extrinsic geometry of quantum states},
  journal       = {Phys. Rev. B},
  volume        = {107},
  number        = {24},
  pages         = {245136},
  year          = {2023},
  doi           = {10.1103/PhysRevB.107.245136},
  url           = {https://doi.org/10.1103/PhysRevB.107.245136},
}

@article{Simon1993,
  author        = {Simon, R. and Mukunda, N.},
  title         = {Bargmann invariant and the geometry of the G\"uoy effect},
  journal       = {Phys. Rev. Lett.},
  volume        = {70},
  number        = {7},
  pages         = {880--883},
  year          = {1993},
  doi           = {10.1103/PhysRevLett.70.880},
  url           = {https://doi.org/10.1103/PhysRevLett.70.880},
}

@article{hassan2018,
  author        = {Hassan, S. R. and Shankar, R. and Chakrabarti, Ankita},
  title         = {Quantum geometry of correlated many-body states},
  journal       = {Phys. Rev. B},
  volume        = {98},
  number        = {23},
  pages         = {235134},
  year          = {2018},
  doi           = {10.1103/PhysRevB.98.235134},
  url           = {https://doi.org/10.1103/PhysRevB.98.235134},
}

@article{bargmann1964,
  author        = {Bargmann, V.},
  title         = {Note on {Wigner}'s {Theorem} on {Symmetry} {Operations}},
  journal       = {J. Math. Phys.},
  volume        = {5},
  number        = {7},
  pages         = {862--868},
  year          = {1964},
  doi           = {10.1063/1.1704188},
  url           = {https://doi.org/10.1063/1.1704188},
}

@misc{huang2025gaussbonnet,
  author        = {Shin-Ming Huang},
  title         = {A Gauss-Bonnet Theorem for Quantum States: Gauss Curvature and Topology in the Projective Hilbert Space},
  year          = {2025},
  eprint        = {2510.15760},
  archivePrefix = {arXiv},
  primaryClass  = {quant-ph},
  url           = {https://arxiv.org/abs/2510.15760},
}

@article{graf2021,
  author        = {Graf, Ansgar and Pi\'echon, Fr\'ed\'eric},
  title         = {Berry curvature and quantum metric in $N$-band systems: An eigenprojector approach},
  journal       = {Phys. Rev. B},
  volume        = {104},
  number        = {8},
  pages         = {085114},
  year          = {2021},
  doi           = {10.1103/PhysRevB.104.085114},
  url           = {https://doi.org/10.1103/PhysRevB.104.085114},
}

@article{wilczek1984,
  author        = {Wilczek, Frank and Zee, A.},
  title         = {Appearance of Gauge Structure in Simple Dynamical Systems},
  journal       = {Phys. Rev. Lett.},
  volume        = {52},
  pages         = {2111--2114},
  year          = {1984},
  doi           = {10.1103/PhysRevLett.52.2111},
  url           = {https://doi.org/10.1103/PhysRevLett.52.2111},
}

@misc{bouhon2023,
  author        = {Adrien Bouhon and Abigail Timmel and Robert-Jan Slager},
  title         = {Quantum geometry beyond projective single bands},
  year          = {2023},
  eprint        = {2303.02180},
  archivePrefix = {arXiv},
  primaryClass  = {cond-mat.mes-hall},
  url           = {https://arxiv.org/abs/2303.02180},
}

@article{mitscherling2025,
  author        = {Mitscherling, Johannes and Avdoshkin, Alexander and Moore, Joel E.},
  title         = {Gauge-invariant projector calculus for quantum state geometry and applications to observables in crystals},
  journal       = {Phys. Rev. B},
  volume        = {112},
  pages         = {085104},
  year          = {2025},
  doi           = {10.1103/qscv-qxqt},
  url           = {https://doi.org/10.1103/qscv-qxqt},
}

@article{Bendokat2024,
  author        = {Bendokat, Thomas and Zimmermann, Ralf and Absil, P.-A.},
  title         = {A {Grassmann} manifold handbook: Basic geometry and computational aspects},
  journal       = {Adv. Comput. Math.},
  volume        = {50},
  pages         = {6},
  year          = {2024},
  doi           = {10.1007/s10444-023-10090-8},
  url           = {https://doi.org/10.1007/s10444-023-10090-8},
}

@book{bernevig2013,
  author        = {Bernevig, B Andrei},
  title         = {Topological insulators and topological superconductors},
  publisher     = {Princeton university press},
  year          = {2013},
}

@article{provost1980,
  author        = {Provost, J. P. and Vall\'ee, G.},
  title         = {Riemannian structure on manifolds of quantum states},
  journal       = {Commun. Math. Phys.},
  volume        = {76},
  pages         = {289--301},
  year          = {1980},
  doi           = {10.1007/BF02193559},
  url           = {https://doi.org/10.1007/BF02193559},
}

@article{Simon1983,
  author        = {Simon, Barry},
  title         = {Holonomy, the Quantum Adiabatic Theorem, and {Berry}'s Phase},
  journal       = {Phys. Rev. Lett.},
  volume        = {51},
  pages         = {2167--2170},
  year          = {1983},
  doi           = {10.1103/PhysRevLett.51.2167},
  url           = {https://doi.org/10.1103/PhysRevLett.51.2167},
}

@article{berry1984,
  author        = {Berry, M. V.},
  title         = {Quantal phase factors accompanying adiabatic changes},
  journal       = {Proc. R. Soc. Lond. A},
  volume        = {392},
  pages         = {45--57},
  year          = {1984},
  doi           = {10.1098/rspa.1984.0023},
  url           = {https://doi.org/10.1098/rspa.1984.0023},
}

@article{Ma2010,
  author        = {Ma, Yu-Quan and Chen, Shu and Fan, Heng and Liu, Wu-Ming},
  title         = {Abelian and non-{Abelian} quantum geometric tensor},
  journal       = {Phys. Rev. B},
  volume        = {81},
  pages         = {245129},
  year          = {2010},
  doi           = {10.1103/PhysRevB.81.245129},
  url           = {https://doi.org/10.1103/PhysRevB.81.245129},
}

@article{Torma2023,
  author        = {T{\"o}rm{\"a}, P{\"a}ivi},
  title         = {Essay: Where Can Quantum Geometry Lead Us?},
  journal       = {Phys. Rev. Lett.},
  volume        = {131},
  pages         = {240001},
  year          = {2023},
  doi           = {10.1103/PhysRevLett.131.240001},
  url           = {https://doi.org/10.1103/PhysRevLett.131.240001},
}

@article{OnishiFu2024,
  author        = {Onishi, Yugo and Fu, Liang},
  title         = {Fundamental Bound on Topological Gap},
  journal       = {Phys. Rev. X},
  volume        = {14},
  pages         = {011052},
  year          = {2024},
  doi           = {10.1103/PhysRevX.14.011052},
  url           = {https://doi.org/10.1103/PhysRevX.14.011052},
}

@article{brouder2007,
  author        = {Brouder, Christian and Panati, Gianluca and Calandra, Matteo and Mourougane, Christophe and Marzari, Nicola},
  title         = {Exponential Localization of {Wannier} Functions in Insulators},
  journal       = {Phys. Rev. Lett.},
  volume        = {98},
  pages         = {046402},
  year          = {2007},
  doi           = {10.1103/PhysRevLett.98.046402},
  url           = {https://doi.org/10.1103/PhysRevLett.98.046402},
}

@article{Ahn2022,
  author        = {Ahn, Junyeong and Guo, Guang-Yu and Nagaosa, Naoto and Vishwanath, Ashvin},
  title         = {Riemannian geometry of resonant optical responses},
  journal       = {Nat. Phys.},
  volume        = {18},
  pages         = {290--295},
  year          = {2022},
  doi           = {10.1038/s41567-021-01465-z},
  url           = {https://doi.org/10.1038/s41567-021-01465-z},
}

@article{Gao2023,
  author        = {Gao, Anyuan and Liu, Yu-Fei and Qiu, Jian-Xiang and Ghosh, Barun and Trevisan, Tha\'is V. and Onishi, Yugo and others},
  title         = {Quantum metric nonlinear {Hall} effect in a topological antiferromagnetic heterostructure},
  journal       = {Science},
  volume        = {381},
  pages         = {181--186},
  year          = {2023},
  doi           = {10.1126/science.adf1506},
  url           = {https://doi.org/10.1126/science.adf1506},
}

@article{Wang2023,
  author        = {Wang, Naizhou and Kaplan, Daniel and Zhang, Zhaowei and Holder, Tobias and Cao, Ning and Wang, Aifeng and Zhou, Xiaoyuan and Zhou, Feifei and Jiang, Zhengzhi and Zhang, Chusheng and Ru, Shihao and Cai, Hongbing and Watanabe, Kenji and Taniguchi, Takashi and Yan, Binghai and Gao, Weibo},
  title         = {Quantum-metric-induced nonlinear transport in a topological antiferromagnet},
  journal       = {Nature},
  volume        = {621},
  pages         = {487--492},
  year          = {2023},
  doi           = {10.1038/s41586-023-06363-3},
  url           = {https://doi.org/10.1038/s41586-023-06363-3},
}

@article{Cuerda2024,
  author        = {Cuerda, Javier and Taskinen, Jani M. and K\"allman, Nicki and Grabitz, Leo and T\"orm\"a, P\"aivi},
  title         = {Observation of quantum metric and non-{Hermitian} {Berry} curvature in a plasmonic lattice},
  journal       = {Phys. Rev. Research},
  volume        = {6},
  pages         = {L022020},
  year          = {2024},
  doi           = {10.1103/PhysRevResearch.6.L022020},
  url           = {https://doi.org/10.1103/PhysRevResearch.6.L022020},
}

@article{Jiang2025,
  author        = {Jiang, Yiyang and Holder, Tobias and Yan, Binghai},
  title         = {Revealing quantum geometry in nonlinear quantum materials},
  journal       = {Rep. Prog. Phys.},
  volume        = {88},
  number        = {7},
  year          = {2025},
  doi           = {10.1088/1361-6633/ade454},
  url           = {https://doi.org/10.1088/1361-6633/ade454},
}

@article{Kim2025Science,
  author        = {Kim, Sunje and Chung, Yoonah and Qian, Yuting and Park, Soobin and Jozwiak, Chris and Rotenberg, Eli and Bostwick, Aaron and Kim, Keun Su and Yang, Bohm-Jung},
  title         = {Direct measurement of the quantum metric tensor in solids},
  journal       = {Science},
  volume        = {388},
  pages         = {1050--1054},
  year          = {2025},
  doi           = {10.1126/science.ado6049},
  url           = {https://doi.org/10.1126/science.ado6049},
}

@article{Sala2025Science,
  author        = {Sala, Giacomo and Mercaldo, Maria Teresa and Domi, Klevis and Gariglio, Stefano and Cuoco, Mario and Ortix, Carmine and Caviglia, Andrea D.},
  title         = {The quantum metric of electrons with spin-momentum locking},
  journal       = {Science},
  volume        = {389},
  pages         = {822--825},
  year          = {2025},
  doi           = {10.1126/science.adq3255},
  url           = {https://doi.org/10.1126/science.adq3255},
}

@article{Sala2026NatMater,
  author        = {Sala, Giacomo and Longo, Emanuele and Mercaldo, Maria Teresa and Gariglio, Stefano and Cuoco, Mario and Mantovan, Roberto and Ortix, Carmine and Caviglia, Andrea D.},
  title         = {Probing the quantum metric of {3D} topological insulators},
  journal       = {Nat. Mater.},
  year          = {2026},
  doi           = {10.1038/s41563-026-02617-3},
  url           = {https://doi.org/10.1038/s41563-026-02617-3},
}

@misc{Avdoshkin2024,
  author        = {Avdoshkin, Alexander},
  title         = {Geometry of degenerate quantum states, configurations of $m$-planes and invariants on complex {Grassmannians}},
  year          = {2024},
  eprint        = {2404.03234},
  archivePrefix = {arXiv},
  primaryClass  = {quant-ph},
  doi           = {10.48550/arXiv.2404.03234},
  url           = {https://doi.org/10.48550/arXiv.2404.03234},
}

@article{Zhang2025,
  author        = {Zhang, Lin and Xie, Bing and Li, Bo},
  title         = {Geometry of sets of {Bargmann} invariants},
  journal       = {Phys. Rev. A},
  volume        = {111},
  pages         = {042417},
  year          = {2025},
  doi           = {10.1103/PhysRevA.111.042417},
  url           = {https://doi.org/10.1103/PhysRevA.111.042417},
}

@article{Yu2025Wilson,
  author        = {Yu, Jiabin and Herzog-Arbeitman, Jonah and Bernevig, B. Andrei},
  title         = {Universal {Wilson} Loop Bound of Quantum Geometry},
  journal       = {Phys. Rev. Lett.},
  volume        = {135},
  pages         = {086401},
  year          = {2025},
  doi           = {10.1103/mp2c-zzkt},
  url           = {https://doi.org/10.1103/mp2c-zzkt},
}

@article{YuReview2025,
  author        = {Yu, Jiabin and Bernevig, B. Andrei and Queiroz, Raquel and Rossi, Enrico and T{\"o}rm{\"a}, P{\"a}ivi and Yang, Bohm-Jung},
  title         = {Quantum geometry in quantum materials},
  journal       = {npj Quantum Mater.},
  volume        = {10},
  pages         = {101},
  year          = {2025},
  doi           = {10.1038/s41535-025-00801-3},
  url           = {https://doi.org/10.1038/s41535-025-00801-3},
}

@article{Kang2025,
  author        = {Kang, Mingu and Kim, Sunje and Qian, Yuting and Neves, Paul M. and Ye, Linda and Jung, Junseo and Puntel, Denny and Mazzola, Federico and Fang, Shiang and Jozwiak, Chris and Bostwick, Aaron and Rotenberg, Eli and Fuji, Jun and Vobornik, Ivana and Park, Jae-Hoon and Checkelsky, Joseph G. and Yang, Bohm-Jung and Comin, Riccardo},
  title         = {Measurements of the quantum geometric tensor in solids},
  journal       = {Nat. Phys.},
  volume        = {21},
  pages         = {110--117},
  year          = {2025},
  doi           = {10.1038/s41567-024-02678-8},
  url           = {https://doi.org/10.1038/s41567-024-02678-8},
}

@article{Oancea2026,
  author        = {Oancea, Marius A. and Mieling, Thomas B. and Palumbo, Giandomenico},
  title         = {Quantum geometric tensors from sub-bundle geometry},
  journal       = {Quantum},
  volume        = {10},
  pages         = {1965},
  year          = {2026},
  doi           = {10.22331/q-2026-01-14-1965},
  url           = {https://doi.org/10.22331/q-2026-01-14-1965},
}

@article{HuangGiataganas2026,
  author        = {Huang, Shin-Ming and Giataganas, Dimitrios},
  title         = {Exploring {Grassmann} manifolds in topological systems via quantum distance},
  journal       = {Phys. Rev. B},
  year          = {2026},
  note          = {in press},
  doi           = {10.1103/qdv4-79lc},
  url           = {https://doi.org/10.1103/qdv4-79lc},
  eprint        = {2412.20046},
  archivePrefix = {arXiv},
  primaryClass  = {quant-ph},
}

@article{MeraMitscherling2022,
  author        = {Mera, Bruno and Mitscherling, Johannes},
  title         = {Nontrivial quantum geometry of degenerate flat bands},
  journal       = {Phys. Rev. B},
  volume        = {106},
  pages         = {165133},
  year          = {2022},
  doi           = {10.1103/PhysRevB.106.165133},
  url           = {https://doi.org/10.1103/PhysRevB.106.165133},
}

@article{Avdoshkin2025,
  author        = {Avdoshkin, Alexander and Mitscherling, Johannes and Moore, Joel E.},
  title         = {Multistate Geometry of Shift Current and Polarization},
  journal       = {Phys. Rev. Lett.},
  volume        = {135},
  pages         = {066901},
  year          = {2025},
  doi           = {10.1103/w761-8nf7},
  url           = {https://doi.org/10.1103/w761-8nf7},
}

@article{Liu2025GLL,
  author        = {Liu, Zhao and Mera, Bruno and Fujimoto, Manato and Ozawa, Tomoki and Wang, Jie},
  title         = {Theory of Generalized Landau Levels and Its Implications for Non-{Abelian} States},
  journal       = {Phys. Rev. X},
  volume        = {15},
  pages         = {031019},
  year          = {2025},
  doi           = {10.1103/1zg9-qbd6},
  url           = {https://doi.org/10.1103/1zg9-qbd6},
}

@article{Ulrich2026,
  author        = {Ulrich, Yannis and Mitscherling, Johannes and Classen, Laura and Schnyder, Andreas P.},
  title         = {Quantum geometric origin of the intrinsic nonlinear {Hall} effect},
  journal       = {Phys. Rev. B},
  volume        = {113},
  pages         = {L201107},
  year          = {2026},
  doi           = {10.1103/4z8z-4kch},
  url           = {https://doi.org/10.1103/4z8z-4kch},
}

@inbook{Neupert_2018,
  author        = {Neupert, Titus and Schindler, Frank},
  title         = {Topological Crystalline Insulators},
  booktitle     = {Topological Matter},
  publisher     = {Springer International Publishing},
  pages         = {31--61},
  year          = {2018},
  isbn          = {9783319763880},
  doi           = {10.1007/978-3-319-76388-0_2},
  url           = {https://doi.org/10.1007/978-3-319-76388-0_2},
}

@article{Bradlyn2022,
  author        = {Bradlyn, Barry and Iraola, Mikel},
  title         = {Lecture notes on Berry phases and topology},
  journal       = {SciPost Phys. Lect. Notes},
  pages         = {51},
  year          = {2022},
  doi           = {10.21468/SciPostPhysLectNotes.51},
  url           = {https://doi.org/10.21468/SciPostPhysLectNotes.51},
}

@article{Moore2007,
  author        = {Moore, J. E. and Balents, L.},
  title         = {Topological invariants of time-reversal-invariant band structures},
  journal       = {Phys. Rev. B},
  volume        = {75},
  number        = {12},
  pages         = {121306(R)},
  year          = {2007},
  doi           = {10.1103/PhysRevB.75.121306},
  url           = {https://doi.org/10.1103/PhysRevB.75.121306},
}

@article{bradlyn2017,
  author        = {Bradlyn, Barry and Elcoro, Luis and Cano, Jennifer and Vergniory, Maia G and Wang, Zhijun and Felser, Claudia and Aroyo, Mois I and Bernevig, B Andrei},
  title         = {Topological quantum chemistry},
  journal       = {Nature},
  volume        = {547},
  number        = {7663},
  pages         = {298--305},
  year          = {2017},
  doi           = {10.1038/nature23268},
  url           = {https://doi.org/10.1038/nature23268},
}

@article{bellissard1994,
  author        = {Bellissard, Jean and van Elst, Andreas and Schulz-Baldes, Hermann},
  title         = {The noncommutative geometry of the quantum Hall effect},
  journal       = {J. Math. Phys.},
  volume        = {35},
  number        = {10},
  pages         = {5373--5451},
  year          = {1994},
  doi           = {10.1063/1.530758},
  url           = {https://doi.org/10.1063/1.530758},
}

@article{Lin2023,
  author        = {Lin, Ling and Ke, Yongguan and Zhang, Li and Lee, Chaohong},
  title         = {Calculations of the Chern number: Equivalence of real-space and twisted-boundary-condition formulas},
  journal       = {Phys. Rev. B},
  volume        = {108},
  number        = {17},
  pages         = {174204},
  year          = {2023},
  doi           = {10.1103/PhysRevB.108.174204},
  url           = {https://doi.org/10.1103/PhysRevB.108.174204},
}

@article{Shiina2025,
  author        = {Shiina, Tsuyoshi and Hamano, Fumina and Fukui, Takahiro},
  title         = {Real-space representation of the second Chern number},
  journal       = {Phys. Rev. B},
  volume        = {111},
  number        = {24},
  pages         = {245135},
  year          = {2025},
  doi           = {10.1103/ztk4-ckj1},
  url           = {https://doi.org/10.1103/ztk4-ckj1},
}

@article{avron1983homotopy,
  author        = {Avron, Joseph E and Seiler, Ruedi and Simon, Barry},
  title         = {Homotopy and quantization in condensed matter physics},
  journal       = {Phys. Rev. Lett.},
  volume        = {51},
  number        = {1},
  pages         = {51},
  year          = {1983},
  doi           = {10.1103/PhysRevLett.51.51},
  url           = {https://doi.org/10.1103/PhysRevLett.51.51},
}

@book{nakahara2018geometry,
  author        = {Nakahara, Mikio},
  title         = {Geometry, topology and physics},
  publisher     = {CRC press},
  year          = {2018},
  doi           = {10.1201/9781315275826},
  url           = {https://doi.org/10.1201/9781315275826},
}

@article{avron1989chern,
  author        = {Avron, J. E. and Sadun, L. and Segert, J. and Simon, B.},
  title         = {Chern numbers, quaternions, and {Berry}'s phases in {Fermi} systems},
  journal       = {Commun. Math. Phys.},
  volume        = {124},
  number        = {4},
  pages         = {595--627},
  year          = {1989},
  doi           = {10.1007/BF01218452},
  url           = {https://doi.org/10.1007/BF01218452},
}

@article{Mera2021_Kahler,
  author        = {Mera, Bruno and Ozawa, Tomoki},
  title         = {K\"ahler geometry and Chern insulators: Relations between topology and the quantum metric},
  journal       = {Phys. Rev. B},
  volume        = {104},
  number        = {4},
  pages         = {045104},
  year          = {2021},
  doi           = {10.1103/PhysRevB.104.045104},
  url           = {https://doi.org/10.1103/PhysRevB.104.045104},
}

@article{Mera2021_engineering,
  author        = {Mera, Bruno and Ozawa, Tomoki},
  title         = {Engineering geometrically flat Chern bands with Fubini-Study K\"ahler structure},
  journal       = {Phys. Rev. B},
  volume        = {104},
  number        = {11},
  pages         = {115160},
  year          = {2021},
  doi           = {10.1103/PhysRevB.104.115160},
  url           = {https://doi.org/10.1103/PhysRevB.104.115160},
}

@article{Schnyder2008,
  author        = {Schnyder, Andreas P. and Ryu, Shinsei and Furusaki, Akira and Ludwig, Andreas W. W.},
  title         = {Classification of topological insulators and superconductors in three spatial dimensions},
  journal       = {Phys. Rev. B},
  volume        = {78},
  number        = {19},
  pages         = {195125},
  year          = {2008},
  doi           = {10.1103/PhysRevB.78.195125},
  url           = {https://doi.org/10.1103/PhysRevB.78.195125},
}
\end{document}